\documentclass[11pt]{article}

\usepackage{booktabs}
\usepackage{amsmath}
\usepackage{amssymb}
\usepackage{hhline}
\usepackage[scale={.75,.75}]{geometry}
\usepackage{url}
\usepackage{caption}
\usepackage[numbers, sort&compress]{natbib}
\usepackage{epsfig}
\usepackage{multirow}

\newcommand{\GeV}{\ \text{GeV}}
\newcommand{\sr}{\ \text{sr}}

\newcommand{\TeV}{\ \text{TeV}}

\newcommand{\kpc}{\ \text{kpc}}
\newcommand{\keV}{\ \text{keV}}
\newcommand{\MeV}{\ \text{MeV}}
\newcommand{\s}{\ \text{s}}
\newcommand{\cm}{\ \text{cm}}
\newcommand{\fex}{\textit{e.g.}~}
\newcommand{\cp}{\textit{cp.}~}

\usepackage{color}
\newcommand{\MP}{M_{\text{P}}}
\newcommand{\GF}{G_{\text{F}}}
\newcommand{\mG}{m_{3/2}}
\newcommand{\mN}{m_{\chi_1^0}}
\newcommand{\mstau}{m_{\tilde \tau_1}}

\begin{document}
\date{\mbox{ }}

\title{ 
{\normalsize     
DESY 11-155\hfill\mbox{}\\
MPP-2011-111\hfill\mbox{}\\}
\vspace{1cm}
\bf \boldmath 
Probing Dark Matter Decay and Annihilation with Fermi LAT Observations of
Nearby Galaxy Clusters
\\[8mm]}

\author{Xiaoyuan~Huang$^{a,b}$,
Gilles~Vertongen$^{c,d}$ and Christoph~Weniger$^{b}$\footnote{Email
addresses: \texttt{huang@mppmu.mpg.de} (X.~Huang),
\texttt{gilles.vertongen@desy.de} (G.~Vertongen) and
\texttt{weniger@mppmu.mpg.de} (C.~Weniger)} \\[8mm]
{\normalsize \it$^a$ National Astronomical Observatories, Chinese Academy of Sciences, Beijing, 100012, China}\\
{\normalsize \it$^b$ Max-Planck-Institut f\"ur Physik, F\"ohringer Ring 6, 80805 M\"unchen, Germany}\\
{\normalsize \it $^c$ Deutsches Elektronen-Synchrotron (DESY),
 \it Notkestrasse 85, 22603 Hamburg, Germany}\\
{\normalsize \it $^d$ Institut d'Astrophysique de Paris, UMR-7095 du CNRS, 98
bis bd Arago, 75014 Paris, France}
}
\maketitle

\thispagestyle{empty}

\begin{abstract}
  \noindent
  Galaxy clusters are promising targets for indirect dark matter searches.
  Gamma-ray signatures from the decay or annihilation of dark matter particles
  inside these clusters could be observable with the Fermi Large Area
  Telescope (LAT). Based on three years of Fermi LAT gamma-ray data, we
  analyze the flux coming from eight nearby clusters individually as well as
  in a combined likelihood analysis.  Concentrating mostly on signals from
  dark matter decay, we take into account uncertainties of the cluster masses
  as determined by X-ray observations and model the cluster emission as
  extended sources. Searching for different hadronic and leptonic decay and
  annihilation spectra, we do not find significant emission from any of the
  considered clusters and present limits on the dark matter lifetime and
  annihilation cross-section.  We compare our lifetime limits derived from
  cluster observations with the limits that can be obtained from the
  extragalactic gamma-ray background (EGBG), and find that in case of hadronic
  decay the cluster limits become competitive at dark matter masses below a
  few hundred GeV. In case of leptonic decay, however, galaxy cluster
  limits are stronger than the limits from the EGBG over the full considered
  mass range.  Finally, we show that in presence of dark matter substructures
  down to $10^{-6}$ solar masses the limits on the dark matter annihilation
  cross-section could improve by a factor of a few hundred, possibly going
  down to the thermal cross-section of $3\times10^{-26}\cm^3s^{-1}$ for dark
  matter masses $\lesssim150\GeV$ and annihilation into $b\bar{b}$. As a
  direct application of our results, we derive limits on the lifetime of
  gravitino dark matter in scenarios with $R$-parity violation.  Implications
  of these limits for the possible observation of long-lived superparticles at
  the LHC are discussed.
\end{abstract}

\newpage

\section{Introduction}
Galaxy clusters are the most massive gravitationally collapsed objects in the
Universe. Historically, the kinematical study of the Coma cluster provided the
first indication for the existence of dark matter~\cite{Zwicky:1933}. Further
gravitational evidence for dark matter was found at many places, \fex~in
stellar rotation curves of spiral galaxies or as baryon acoustic oscillations
imprinted in the cosmic microwave background (for reviews on particle dark
matter see Refs.~\cite{Jungman:1995df, Bergstrom:2000pn, Bertone:2004pz}). By
now, the $\Lambda$CDM scenario is the standard framework for cosmology,
leaving open the question of what the nature of the dark matter particles is.
Theoretical models for dark matter predict a large variety of possible
non-gravitational signatures that would help to reveal its properties.
However, despite lots of efforts, none of these signals has been unambiguously
detected so far.

If dark matter is made out of WIMPs (Weakly Interacting Massive Particles),
their efficient self-annihilation in the early Universe would explain the
observed dark matter density. Today, the same annihilation process could
contribute to the measured cosmic-ray fluxes; a clear detection of the
annihilation products would reveal information about the dark matter
particle's mass and interactions. Similar signatures could be produced if dark
matter is unstable and decays, providing us with information on the lifetime
of the dark matter particle. Indirect searches for dark matter are aiming at
seeing such annihilation or decay signals above the astrophysical backgrounds.
These searches typically concentrate on photons or neutrinos, which carry
spatial and spectral information about their origin, and on different
anti-matter species with their relatively low astrophysical backgrounds.
Together with satellite galaxies of the Milky Way~\cite{Essig:2009jx,
Scott:2009jn, :2010zzt, Abdo:2010ex, GeringerSameth:2011iw, Aleksic:2011jx,
collaboration:2011wa} and the Galactic center~\cite{Aharonian:2006wh,
Dobler:2009xz, Hooper:2010mq, Dobler:2011mk, Ellis:2011du, Vitale:2011zz,
Abramowski:2011hc, Bringmann:2011ye}, nearby massive galaxy clusters are among
the most promising targets for indirect dark matter searches by means of gamma
rays~\cite{Colafrancesco:2005ji, Jeltema:2008vu, Baltz:2008wd,
Aharonian:2009bc, Aleksic:2009ir, Pinzke:2009cp, Ackermann:2010rg,
Dugger:2010ys, Ackermann:2010qj, SanchezConde:2011ap, Pinzke:2011ek}.\medskip

A theoretically well motivated example for \textit{decaying dark matter} is
the gravitino $\psi_{3/2}$, which appears in locally supersymmetric extensions
of the Standard Model. In scenarios where $R$-parity is mildly violated and
the gravitino is the lightest superparticle (LSP), thermal leptogenesis,
gravitino dark matter and primordial nucleosynthesis are naturally
consistent~\cite{Buchmuller:2007ui}.  Within this framework, the gravitino
would decay with cosmological lifetimes~\cite{Takayama:2000uz}, making its
decay products potentially observable in the cosmic-ray
fluxes~\cite{Bertone:2007aw, Ibarra:2007wg, Lola:2007rw, Ibarra:2008qg,
Ishiwata:2008cu, Ishiwata:2008tp, Covi:2008jy, Buchmuller:2009xv, Choi:2009ng,
Bomark:2009zm, Choi:2010xn, Choi:2010jt}.  For gravitino masses $\lesssim
100\GeV$, the most prominent feature in the decay spectrum is an intense
gamma-ray line, produced by the two-body decay into neutrinos and photons,
$\psi_{3/2}\to\gamma\nu$~\cite{Ibarra:2007wg}.  Dedicated searches for such a
feature in the current gamma-ray observations of the Fermi
LAT~\cite{Atwood:2009ez} exist in the literature, see Refs.~\cite{Abdo:2010nc,
Vertongen:2011mu}, and their null results were used to put lower limits on the
gravitino lifetime around $6\times 10^{28}\s$~\cite{Vertongen:2011mu}.
However, for larger gravitino masses $\gtrsim 100\GeV$, the branching ratio
into gamma-ray lines is strongly suppressed, and instead decay modes like
$\psi_{3/2}\to W^\pm \ell^\mp$ and $\psi_{3/2}\to Z^0\nu$ produce a gamma-ray
flux with a broad continuous energy spectrum. It is this flux that could
potentially show up in observations of galaxy clusters, whereas the
observation of the gamma-ray line in galaxy clusters would be difficult due to
the limited statistics.

In general, dark matter lifetimes of the order of $10^{26}$--$10^{29}$s, which
is in the ballpark of what is accessible experimentally, are obtained when the
symmetry responsible for the dark matter stability is violated by dimension
six operators generated close to the grand unification
scale~\cite{Eichler:1989br, Arvanitaki:2008hq, Arina:2009uq}. Indeed, models
of this kind were proposed to explain the $e^\pm$ ``excesses'' observed by the
PAMELA~\cite{Adriani:2008zr}, Fermi~LAT~\cite{Abdo:2009zk, Ackermann:2010ij}
and H.E.S.S.~\cite{Aharonian:2008aa, Aharonian:2009ah} experiments. To avoid
the stringent anti-proton limits~\cite{Adriani:2008zr, Adriani:2010rc}, the
decay should be mostly leptophilic (see \fex Refs.~\cite{Fox:2008kb,
Kyae:2009jt, Ibarra:2009bm, Bi:2009uj, Spolyar:2009kx, Cohen:2009fz,
Chun:2009zx}), and one typical decay mode that could well reproduce the
locally observed $e^\pm$ fluxes is the decay into muons, $\psi\to\mu^+\mu^-$,
with a large dark matter mass around $m_\psi\simeq3\TeV$ and lifetimes around
$2\times10^{26}\s$~\cite{Ibarra:2009dr, Meade:2009iu}.  Inside galaxy
clusters, due to inverse Compton scattering (ICS) on the cosmic microwave
background (CMB), almost all of the kinetic energy of the produced
high-energetic electrons and positrons is transferred into gamma rays, with
energies up to $\mathcal{O}(100\GeV)$. This makes possible the investigation
of the decaying and annihilating dark matter interpretations of the $e^\pm$
excesses by galaxy cluster observations.  \bigskip

A dedicated search for dark matter annihilation signals from galaxy clusters,
using 11 months of Fermi LAT data, was carried out in
Ref.~\cite{Ackermann:2010rg}. The null result of this search was used to
derive limits on the dark matter annihilation rate into $b\bar{b}$ and
$\mu^-\mu^+$.  In Ref.~\cite{Dugger:2010ys} these results were translated into
limits on the dark matter decay rate, and it was demonstrated that galaxy
cluster observations give strong constraints on the dark matter lifetime,
superior to the limits that could be obtained from satellite galaxy
observations, and of the order of the limits that can be derived from the
extragalactic gamma-ray background. This makes galaxy clusters promising
targets when searching for signals from dark matter decay. Concerning WIMP
dark matter, taking into account the expected boost of the annihilation signal
due to dark matter substructures in the cluster halo, limits can potentially
go down to the cross-section expected from thermal
freeze-out~\cite{SanchezConde:2011ap, Pinzke:2011ek}. Further studies of the
galaxy cluster emission as seen by the Fermi LAT, H.E.S.S. and MAGIC were
presented in Ref.~\cite{Ackermann:2010qj, Aharonian:2009bc, Aleksic:2009ir},
some applications to annihilating and decaying dark matter models were
discussed in Refs.~\cite{Yuan:2010gn, Ke:2011xw, Luo:2011bn}. 

Besides the large amount of dark matter, it is known from radio observations
that galaxy clusters are also a host for energetic cosmic rays, which can be
accelerated during the process of cluster formation by mergers or accretion
shocks. Proton-proton collision as well as the ICS of an energetic electron
population can produce a possibly observable gamma-ray flux (see \fex
Refs.~\cite{Ackermann:2010rg, Pinzke:2011ek}). Such a flux should however be
finally distinguishable from a dark matter signal through the analysis of the
energy spectra if the statistics is high enough~\cite{Jeltema:2008vu}.

\bigskip

In this paper, we analyze the gamma-ray flux from eight galaxy clusters as
measured by the Fermi LAT since Aug 2008, and we present constraints on the
dark matter lifetime and annihilation cross-section.  We analyze the different
target clusters individually as well as in a combined likelihood approach, and
search for significant gamma-ray emission as an indication for decaying or
annihilating dark matter.  In Ref.~\cite{Ackermann:2010rg, Dugger:2010ys} the
dark matter signal was approximated to be point-source like. Importantly, this
approximation becomes problematic in the case of dark matter decay or
substructure-boosted annihilation, since the extend of the expected signal
starts to exceed the angular resolution of the Fermi LAT. To account for this,
we model the dark matter emission as an extended source.  In absence of a
clear signal, we derive limits on the dark matter lifetime and annihilation
cross-section as function of the dark matter mass, for different hadronic and
leptonic final states. Cluster masses and the expected decay or annihilation
signals are derived from the extended HIFLUGCS catalog~\cite{Reiprich:2001zv,
Chen:2007sz} which is based on ROSAT PSPC X-ray
observations~\cite{Voges:1999ju}, and the corresponding uncertainties are
consistently taken into account.  We compare the obtained lifetime limits with
the limits that can be derived from the extragalactic gamma-ray background,
and we discuss the implications of our limits on the decaying or annihilating
dark matter interpretation of the $e^\pm$ excesses.  Furthermore, we will
illustrate how the limits improve when a boost of the annihilating signal due
to substructures is included. Finally, we apply our findings to the scenario
of decaying gravitino dark matter, and derive new constraints on the gravitino
lifetime for masses above about $100\GeV$.  We comment on implications for the
possible observation of long-lived superparticles at the LHC.\medskip 

The rest of the paper is organized as follows: In section 2 we present our
galaxy cluster analysis of the Fermi LAT data. We discuss the expected dark
matter signals, our treatment of the LAT data and the details of our
statistical analysis. In section 3, we shortly review how limits on the dark
matter lifetime from the extragalactic gamma-ray background are obtained.  Our
results and their discussion are presented in section 4.  Finally, section 5
is devoted to gravitino dark matter, where we present limits on the gravitino
lifetime for gravitino masses above $100\GeV$, as well as the implied limits
on the decay lengths of next-to-lightest superparticles (NLSP) at particle
colliders.  We conclude in section 6.

Throughout this work we assume a $\Lambda$CDM cosmology with the parameters
$\Omega_\Lambda=0.728$ and $h\equiv H_0/100\,\text{km}\;
\text{s}^{-1}\,\text{Mpc}^{-1}=0.704$~\cite{Komatsu:2010fb}.

\section{Galaxy Cluster Analysis}
The eight galaxy clusters that we consider in this work are summarized in
Tab.~\ref{tab:clusters}. They are selected from the extended HIFLUGCS X-ray
catalog~\cite{Reiprich:2001zv, Chen:2007sz} in order to yield large signals
from dark matter decay, but are also among the best clusters when searching
for signals from dark matter annihilation. Galaxy clusters with potentially
large signals that we disregard are: Ophiuchus, A3627 and 3C129 because they
lie too close to the Galactic plane, Centaurus, M49 and A2877 because of
issues with our adopted model for Galactic diffuse emission,\footnote{The
positions of the Centaurus and M49 clusters unfortunately coincide with sharp
edges in our Galactic diffuse emission model, \texttt{gal\_2yearp7v6\_v0}; the
region near A2877 contains a large number of faint sources that are not part
of the 2FGL. In all three cases the background fits are unreliable, and we
neglect these targets from our analysis. Including them would improve our
overall limits.} and Virgo (M87) and Perseus because of the presence of bright
gamma-ray sources at their center~\cite{Abdo:2009ta, Abdo:2009wt}.

\begin{table}[t]
  \centering
  \begin{tabular}{@{}lccccccccc@{}}
    \toprule
    Cluster&\phantom{}&R.A.&Dec.&z&
    $J^\text{dec.}_{\Delta\Omega}$ &
    $J^\text{ann.}_{\Delta\Omega}$ &
    $\theta_s$ \\[1mm]
    &&&&& 
    [$10^{18}$GeVcm$^{-2}$] &
    [$10^{17}$GeV$^{2}$cm$^{-5}$] & 
    [$^\circ$]\\
    \midrule
    Fornax && 54.67 & -35.31 & 0.0046 & $20.3^{+4.6}_{-6.8}$ & $8.8^{+2.0}_{-2.8}$ & $0.44^{+0.07}_{-0.11}$ \\
    Coma && 194.95 & 27.94 & 0.0232 & $10.7^{+1.8}_{-2.7}$ & $1.3^{+0.20}_{-0.31}$ & $0.23^{+0.02}_{-0.04}$ \\
    A1367 && 176.19 & 19.70 & 0.0216 & $10.6^{+1.3}_{-2.9}$ & $1.4^{+0.15}_{-0.34}$ & $0.23^{+0.02}_{-0.04}$ \\
    A1060 && 159.18 & -27.52 & 0.0114 & $10.2^{+2.0}_{-3.5}$ & $2.2^{+0.38}_{-0.69}$ & $0.24^{+0.03}_{-0.06}$ \\
    AWM7 && 43.62 & 41.58 & 0.0172 & $9.9^{+1.9}_{-3.9}$ & $1.6^{+0.27}_{-0.56}$ & $0.22^{+0.03}_{-0.06}$ \\
    S636 && 157.52 & -35.31 & 0.0116 & $6.8^{+1.5}_{-1.7}$ & $1.5^{+0.29}_{-0.34}$ & $0.18^{+0.03}_{-0.03}$ \\
    NGC4636 && 190.71 & 2.69 & 0.0037 & $6.1^{+0.80}_{-1.7}$ & $3.5^{+0.39}_{-0.85}$ & $0.19^{+0.02}_{-0.04}$ \\
    NGC5813 && 225.30 & 1.70 & 0.0064 & $6.0^{+4.6}_{-4.2}$ & $2.2^{+1.4}_{-1.4}$ & $0.18^{+0.08}_{-0.10}$ \\
    \bottomrule
  \end{tabular}
  \caption{Galaxy clusters considered in this work, with their coordinates
  (equatorial J2000.0) and redshift $z$ from Ref.~\cite{Reiprich:2001zv}. We
  show the integrated surface densities
  $J_{\Delta\Omega}^\text{dec.}\equiv\int_{\Delta\Omega}J^\text{dec.}(\Omega)$
  and $J_{\Delta\Omega}^\text{ann.}\equiv\int_{\Delta\Omega}
  J^\text{ann.}(\Omega)$ of the dark matter signal, obtained inside a region
  of $1^\circ$ radius around the cluster center, as well as the projected
  angle $\theta_s$ of the scale radius $r_s$ of the adopted NFW profile.
  Central values and errors for these parameters are derived from the cluster
  masses in Ref.~\cite{Chen:2007sz}.}
  \label{tab:clusters}
\end{table}

\subsection{Dark Matter Signal}
The gamma-ray flux from dark matter annihilation or decay that is expected to
be seen in galaxy cluster observations factorizes into an astrophysical part,
which contains information about the dark matter distribution
$\rho_\text{dm}$, and a particle-physics part, which is universal for all
observed targets. Assuming a spherical dark matter halo, the astrophysical
factor, $J^\text{dec./ann.}(\theta)$, just depends on the cluster-centric
angle $\theta$ and is given by a line-of-sight integral.  In the case of dark
matter decay, the signal flux reads
\begin{equation}
  \frac{dJ_\text{sig}}{dE\ d\Omega}(\theta) =
  \frac{1}{4\pi \,m_\psi\ \tau_\psi}
  \,\frac{dN_\gamma}{dE} 
  \underbrace{\int_\text{l.o.s.} ds
  \,\rho_\text{dm}\left(s, \Omega\right)
  }_{\equiv
  J^\text{dec.}(\theta)}\;,
  \label{eqn:fluxDDM}
\end{equation}
while in the annihilation case, it is given by
\begin{equation}
  \frac{dJ_\text{sig}}{dE\ d\Omega}(\theta) = \frac{\langle \sigma
  v\rangle}{8\pi \,m_\psi^2}
  \,\frac{dN_\gamma}{dE} 
  \underbrace{\int_\text{l.o.s.} ds
  \,\rho_\text{dm}^2\left(s, \Omega\right)
  }_{\equiv
  J^\text{ann.}(\theta)}\;.
  \label{eqn:fluxADM}
\end{equation}
Here, $m_\psi$ denotes the dark matter mass, while $\tau_\psi$ and $\langle
\sigma v\rangle$ are the dark matter lifetime and total annihilation cross
section, respectively. The energy spectrum of gamma rays produced in the
decay/annihilation is given by $dN_\gamma/dE$.  Note that the energy spectrum
$dN_\gamma/dE=dN^\text{prim}_\gamma/dE+dN^\text{IC}_\gamma/dE$ includes prompt
gamma rays that are directly produced in the decay or annihilation process
(final-state radiation, $\pi^0\to\gamma\gamma$ etc.) as well as the gamma rays
that originate from ICS losses of $e^\pm$ from dark matter on the
intra-cluster radiation field. We calculated the energy spectra of gamma rays
and electrons with the event generator \textsf{Pythia
6.4.19}~\cite{Sjostrand:2006za}, and cross-checked our results with the
analytic expressions presented in Ref.~\cite{Cembranos:2010dm}.

\paragraph{Inverse Compton scattering.} 
Electrons and positrons produced in the decay or annihilation of dark matter
particles inside galaxy clusters suffer inverse Compton and synchrotron losses
when interacting with the intra-cluster radiation field. The dominant
component of this radiation field is in most cases the CMB; other
contributions, which can become relevant close to the cluster center, are the
starlight, dust radiation and the intra-cluster magnetic field (see discussion
in Ref.~\cite{Pinzke:2011ek}). In case of dark matter decay or
substructure-boosted annihilation, the possible impact of these additional
components on our results is small, as we will exemplify below for the Coma
cluster; if not stated otherwise, we consider the CMB only throughout this work.

The average energy spectrum of gamma-rays with energy $E_\gamma$ that are
generated by the inverse Compton scattering of one electron with an initial
energy of $E_0$ is given by
\begin{equation}
    \frac{dN^{\rm IC}_\gamma}{d E_\gamma}=
    \int_0^\infty d\epsilon  
    \int_{m_e}^{E_0} dE_e\; \frac{d\sigma^{\rm
    IC}(E_e,\epsilon)}{dE_\gamma} 
    \frac{f_\text{CMB}(\epsilon)}{b_\text{loss}(E_e)}\;.
  \label{eqn:IC-rate}
\end{equation}
Here, $f_\text{CMB}(\epsilon)$ is the CMB energy spectrum with temperature
$T_\text{CMB}=2.725\ \text{K}$, and $d\sigma^\text{IC}/dE_\gamma$ denotes the
differential cross section of inverse Compton scattering of an electron with
energy $E_e$ when a CMB photon with energy $\epsilon$ is up-scattered to
energies between $E_\gamma$ and $E_\gamma+dE_\gamma$.  Due to the very low
energy of the CMB photons, the center-of-mass energy of the processes we are
interested in is always smaller than the electron mass, which allows us to use
the non-relativistic limit of the Klein-Nishina equation in our
calculations\footnote{This approximation breaks down for electron
energies above $m_e^2/T_\text{CMB}\sim 10^3\TeV$.} (see \fex
Ref.~\cite{Blumenthal:1970gc}):
\begin{equation}
    \frac{d\sigma^{\rm IC}(E_e,\epsilon)}{dE_\gamma}=
    \frac{3}{4}\frac{\sigma_{\rm T}}{\gamma_e^2\, \epsilon}
    \left[2q\ln q + 1 + q - 2q^2 \right]\;,
  \label{eqn:ICrate}
\end{equation}
where $\sigma_{\rm T}=0.67\,{\rm barn}$ is the Compton scattering cross
section in the Thomson limit, $\gamma_e\equiv E_e/m_e$ is the Lorentz factor
of the electron, $m_e=511\keV$ is the electron mass, and $q\equiv
E_\gamma/E_\gamma^\text{max}$ with $E_\gamma^\text{max}\equiv 4\gamma_e^2
\epsilon$. Eq.~\eqref{eqn:ICrate} holds in the range $\epsilon\leq E_\gamma
\leq E_\gamma^\text{max}$, in the limit where down-scattering is neglected.

In the above equation, $b_\text{loss}(E_\text{e})$ is the energy loss rate of
an electron with energy $E_\text{e}$. Typically, the main contribution comes
from ICS on the CMB, but we can also include synchrotron losses on the
intra-cluster magnetic field; $b_\text{loss}=b_\text{ICS}+b_\text{syn}$. In
the non-relativistic limit, the ICS and synchrotron losses read
\begin{equation}
  b_\text{ICS}(E_e)=\frac{4}{3}\sigma_\text{T} \gamma_e^2 \underbrace{\int_0^\infty
  d\epsilon\; \epsilon f_\text{CMB}(\epsilon)}_{\equiv \rho_\text{CMB}}
  \quad\text{and}\quad
  b_\text{syn}(E_e)=\frac{4}{3}\sigma_\text{T} \gamma_e^2 \frac{B^2}{2}\;,
\end{equation}
respectively. In order for the magnetic field $B$ to dominate the CMB energy
density $\rho_\text{CMB}$ (namely, $B^2/2 > \rho_\text{CMB}$) in galaxy
clusters it has to exceed the critical value $B_\text{CMB}=3.2\mu\text{G}$
(assuming redshifts $z\ll1$). The energy spectrum of the dark matter induced
ICS radiation would then scale like $\propto(1+(B/3.2\mu\text{B})^2)^{-1}$.

The energy loss time $\tau_\text{loss}=E_e/\dot{E_e}$ of electrons with
$100\GeV$--$10\TeV$ energies, as relevant for our work, lies in the range of
$10^{-4}$--$10^{-2}\ \text{Gy}$. This is much shorter than the cosmic-ray
relaxation times in galaxy clusters which typically are of order $1$--$10\
\text{Gy}$~\cite{Ensslin:2010zh, Pinzke:2011ek}. The propagation scale
corresponding to $\gtrsim100\GeV$ electrons is expected to be
$\lesssim1\kpc$~\cite{Blasi:1999aj, Lavalle:2009fu}, which for Mpc distances
is well below the angular resolution of the LAT~\cite{Fermi:performance}. We
hence neglect effects of cosmic-ray transport and consider that $e^\pm$ are
loosing all their energy instantaneously where they are produced. In this
limit, the angular profile of the ICS signal is identical to the angular
profile of the prompt radiation. This is opposite to the case of dwarf
galaxies, where due to their proximity propagation effects have to be taken
into account in general, see \fex Ref.~\cite{Abdo:2010ex}.  \medskip

After some algebra, one can finally show that the energy spectrum of ICS
radiation emitted from a single electron with an initial energy $E_e$ is given
by the expression ($B\to0$)
\begin{align}
  \frac{dN^{\rm IC}_\gamma}{d E_\gamma}=\frac{9m_e}{32\gamma_e^3}
  \int_{\frac1{4\gamma_e^2}}^1 dq\ \frac{1}{q^{5/2}}
  \left\{ \frac{92}{525}-\frac23q^{3/2}
  -\frac2{25}q^{5/2} +\frac47q^{7/2}-\frac45q^{5/2}\log q \right\}
  \frac{f_\text{CMB}\left(\frac{E_\gamma}{4\gamma_e^2q}\right)}{\rho_\text{CMB}}\;.
  \label{}
\end{align}
A subsequent convolution with the energy spectrum $dN_e/dE_e$ of electrons and
positrons yields then the ICS contribution to the gamma-ray spectrum,
$dN^\text{IC}_\gamma/dE$, in Eqs.~\eqref{eqn:fluxDDM}~and~\eqref{eqn:fluxADM}.

\subsection{Dark Matter Distribution}
\label{sec:targets}
As discussed above, the expected dark matter signals depend crucially on the
dark matter profiles $\rho(r)_\text{dm}$ of the target galaxy clusters. We
assume throughout this work that the smooth component of the dark matter halo
follows a Navarro-Frenk-White (NFW) profile~\cite{Navarro:1996gj,
Abdo:2010nc},
\begin{equation}
  \rho_\text{dm}(r)=\frac{\rho_s}{r/r_s(1+r/r_s)^2}\;,
  \label{eqn:NFWprofile}
\end{equation}
where the scale radius $r_s$ and the density normalization $\rho_s$ have to be
determined from observations.\footnote{We find that using an Einasto
profile~\cite{Graham:2005xx, Navarro:2008kc} with similar $M_{200}$ and
$r_{200}$ (with $\alpha=0.17$, $r_{-2}=r_s$ and $\rho_{-2}\simeq
\rho_s/4.2$~\cite{Pinzke:2011ek}) leaves the results for decaying dark matter
essentially unchanged, whereas the fluxes from dark matter annihilation as
summarized in Tab.~\ref{tab:clusters} are increased by about $\sim30\%$.} The
cluster mass $M_\Delta$ inside a cluster-centric radius $r_\Delta$ is defined
such that the average density inside $r_\Delta$ equals $\Delta$ times the
critical density of the Universe, $\rho_c$ (typically $\Delta\approx100$--$500$).
To determine the parameters of the NFW profile from $M_\Delta$ and $r_\Delta$,
we adopt the observationally obtained concentration-mass relation from
Ref.~\cite{Buote:2006kx},
\begin{equation}
  c_\text{vir}(M_\text{vir}) = 9 \left( \frac{M_\text{vir}}{10^{14}h^{-1}M_\odot}
  \right)^{-0.172}\;,
  \label{eqn:CMrel}
\end{equation}
where the concentration parameter $c\equiv r_\text{vir}/r_s$ relates the
virial radius $r_\text{vir}$ as defined by $\Delta=\Delta_\text{vir}\simeq98$
(see appendix of Ref.~\cite{Hu:2002we} and references therein) to the scaling
radius $r_s$, and $M_\odot = 2.0\times10^{30}\ \text{kg}$ denotes the solar
mass.\footnote{Varying the prefactor in $c_\text{vir}$ in the range $9\pm2$
leaves the fluxes from decaying dark matter essentially unchanged, whereas the
annihilation fluxes increase or decrease by about 30--40\%. Note that CDM
simulations favor a somewhat smaller concentration of about $\sim6$ at
$M_\text{vir}=10^{14}h^{-1}M_\odot$~\cite{Duffy:2008pz, Klypin:2010qw} .} The
virial radius $r_\text{vir}$ is then related to $r_\Delta$
via~\cite{Hu:2002we}
\begin{equation}
  f(r_s/r_\Delta) = \frac{\Delta}{\Delta_\text{vir}}f(r_s/r_\text{vir})\;,
  \label{eqn:fRel}
\end{equation}
where
\begin{equation}
  f(x)=x^3\left[\ln(1+x^{-1})-(1+x)^{-1}\right]\;,
  \label{}
\end{equation}
and
\begin{equation}
  \frac{M_\Delta}{M_\text{vir}}=\frac{\Delta}{\Delta_\text{vir}}\left(
  \frac{r_\Delta}{r_\text{vir}} \right)^3\;.
  \label{eqn:massRelation}
\end{equation}
With Eqs.~\eqref{eqn:CMrel}, \eqref{eqn:fRel} and~\eqref{eqn:massRelation},
one can find $r_s$ and $\rho_s$ as a function of $M_\Delta$ and $r_\Delta$.
\bigskip 

\begin{figure}[t]
  \begin{center}
    \includegraphics[width=\linewidth]{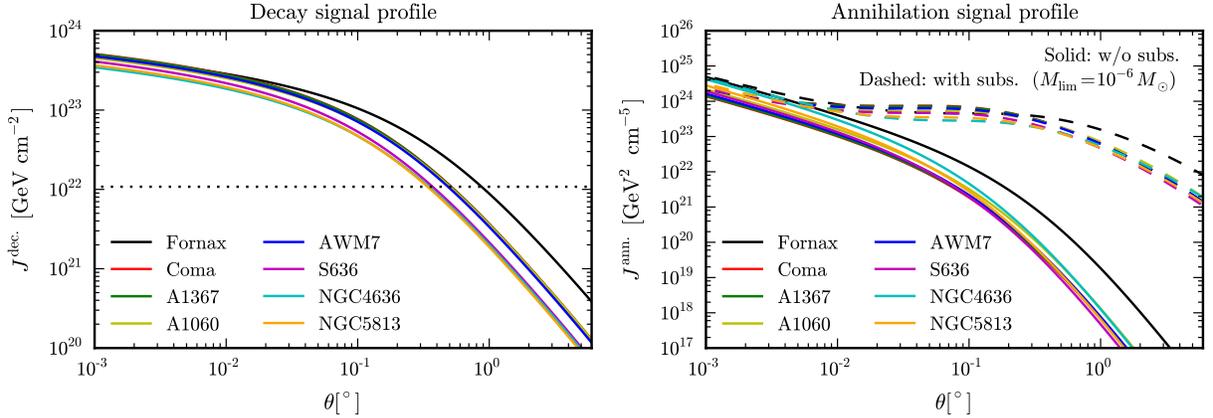}
  \end{center}
  \caption{Left panel: profile of the dark matter decay signal as function of
  the cluster-centric angle $\theta$ (PSF effects not included). The dotted
  line shows the isotropic Galactic contribution to the dark matter signal.
  Right panel: the same for the dark matter annihilation signal. Solid lines
  show the signal coming from the smooth halo component alone, dashed lines
  include effects from dark matter substructures, which boosts the signal at
  angles around $1^\circ$ (as discussed in Section~\ref{sec:boosts}).  For
  comparison: the angular resolution of the LAT (\texttt{P7SOURCE\_V6}) in
  terms of the 68\% containment angle is $6^\circ$ at $100\MeV$, $0.9^\circ$
  at $1\GeV$ and $0.2^\circ$ at $100\GeV$~\cite{Fermi:performance}.}
  \label{fig:profile}
\end{figure}

Using the cluster masses as derived from ROSAT PSPC X-ray observations in the
extended HIFLUGCS catalog~\cite{Chen:2007sz}, we calculate the signal surface
densities $J^\text{ann.}$ and $J^\text{dec.}$ as described above.\footnote{The
values of $M_{500}$ and $r_{500}$ are rescaled to our adopted Hubble constant,
and $M_{500}$ is reduced by the gas fraction $f_\text{gas}$ indicated in
Ref.~\cite{Chen:2007sz}. We checked that using $M_{200}\simeq \sqrt{5/2}\
M_{500}$ and $r_{200}\simeq \sqrt{5/2}\ r_{500}$ as a starting point would
increase the predicted fluxes by 10--20\%.} Our results, as function of the
cluster-centric angle $\theta$, are plotted in Fig.~\ref{fig:profile}. These
profiles are used to model the extended dark matter signal in our analysis.
For convenience and comparison with previous work, we show results for
$J^\text{ann.}$ and $J^\text{dec.}$ integrated over a cluster-centric region
of $1^\circ$ radius in Tab.~\ref{tab:clusters}.  There, we also indicate the
projected scaling angle $\theta_s=r_s/D$, where the distance to the cluster is
given by $D\simeq z c/H_0$.  The signal uncertainties shown in
Tab.~\ref{tab:clusters} are directly derived from the mass uncertainties in
Ref.~\cite{Chen:2007sz} and as large as a factor of two in some cases. Within
the error bars our results agree largely with what was found in
Refs.~\cite{Ackermann:2010rg, Dugger:2010ys} based on the initial HIFLUGCS
catalog~\cite{Reiprich:2001zv}. Besides the uncertainties from the fits to the
X-ray profiles that were already discussed in Ref.~\cite{Reiprich:2001zv}, the
mass ranges quoted in Ref.~\cite{Chen:2007sz} take additionally into account
uncertainties in the X-ray temperature profile, which leads to somewhat larger
error-bars compared to Ref.~\cite{Reiprich:2001zv}.

\subsection{Signal Boost from Dark Matter Substructures} \label{sec:boosts} A
prediction of the cold dark matter paradigm is the hierarchical structuring of
dark matter halos. Dark matter substructures inside of galaxy cluster halos
are observationally known to exist down to the scale of dwarf galaxies,
$10^7M_\odot$; for thermally produced WIMPs they are predicted to continue
down to free streaming masses of about $10^{-6}M_\odot$ and
below~\cite{Hofmann:2001bi, Green:2005fa} (for a discussion of possible ranges
depending on the dark matter model see Ref.~\cite{Bringmann:2009vf}). Since
the dark matter annihilation signal depends on the dark matter density
squared, the existence of substructures can boost the annihilation signal
considerably with respect to the signal from the smooth halo; the details
depend on the mass function of substructures, the concentration mass relation
and the radial distribution (see \fex Ref.~\cite{Pinzke:2011ek} for a recent
discussion).  Dynamical friction and tidal stripping near the cluster center
lead to a local depletion of substructures that results in a relative
enhancement of the boosted signal in the outskirts of the main halo. In
general, the boosted signal is expected to be considerably more extended than
the signal coming from the smooth main halo alone. Deriving the magnitude of
the signal boost relies on extrapolations of numerical simulations for
dissipationless DM~\cite{Springel:2008cc, Springel:2008by, Kuhlen:2008qj} over
many orders of magnitude in the substructure mass. In the literature,
predictions for the substructure boost have not yet converged; in case of
galaxy clusters, signal boosts in the range of
$\sim10$--$50$~\cite{Ackermann:2010rg, SanchezConde:2011ap} up to
$\sim1000$~\cite{Pinzke:2009cp, Pinzke:2011ek} were recently discussed. 
As mentioned above, the actual values strongly depend on the adopted subhalo
mass fraction (which is partially correlated with the value of $\sigma_8$ used
in the underlying simulations), the subhalo mass distribution functions and
the adopted halo concentration. Furthermore, for individual clusters, precise
predictions appear to be difficult since the halo-to-halo scattering of the
substructure fraction, which roughly correlates with the concentration of the
particular halo, can be quite large and
$\mathcal{O}(1)$~\cite{Contini:2011yf}.

In the present paper, we adopt an optimistic scenario and estimate the boosted
dark matter signal following Ref.~\cite{Pinzke:2011ek}: Based on the
high-resolution dissipationless dark matter simulations of the Aquarius
project~\cite{Springel:2008cc} (which features a realtively large subhalo
fraction), the boost of the dark matter annihilation signal was determined  in
Ref.~\cite{Springel:2008by} for a Milky Way sized halo.  Extrapolating the
mass of the smallest subhalos down to $M_\text{lim}=10^{-6}M_\odot$, an
increase in the overall luminosity $L=\int dV \rho_\text{dm}^2$ of about 230
was found, being mostly due to a signal enhancement at large galactocentric
distances.  The luminosity due to substructures inside a radius $r$ is well
fitted by~\cite{Pinzke:2011ek}
\begin{align}
  \label{eqn:L}
  L_\text{sub}(<r) &=a_0 C(M_{200})\ L_{200\rm sm}(M_{200})\  x^{f(x)}\;,\\
  \nonumber
  f(x)&=a_1 x^{a_2}\;,\\
  C(M_{200})&=0.023\left( \frac{M_{200}}{M_\text{lim}} \right)^{\alpha_C}\;,
  \nonumber
\end{align}
where $x\equiv r/r_{200}$, $\alpha_0=0.76$, $\alpha_1=0.95$, $\alpha_2=-0.27$
and $\alpha_C=0.226$. Here, $L_\text{200sm}$ denotes the luminosity of the
smooth halo component inside $r_{200}$ alone. The only free parameter is the
cutoff scale for the dark matter subhalo mass, which we fix to
$M_\text{lim}=10^{-6}M_\odot$. The parameters $M_{200}$ and $r_{200}$ are
directly determined from the adopted NFW profile for each cluster.  The
overall boosted dark matter signal can then be calculated from
Eq.~\eqref{eqn:L}. 

Our results for the signal profile from dark matter annihilation in presence
of substructures are shown in the right panel of Fig.~\ref{fig:profile} by
dashed lines, the signal from the smooth halo alone is shown by solid lines.
In presence of substructures, the annihilation signal extends to radii of
around $1^\circ$, below $0.01^\circ$ it is still dominated by the smooth dark
matter halo. The boost factors that we obtain for the different considered
galaxy clusters inside an opening angle of $\theta_{200}=r_{200}/D$ are in the
range 500--1200, in agreement with Ref.~\cite{Pinzke:2011ek}.  Similar large
values were recently also found in Ref.~\cite{Gao:2011rf} (however, see
discussion above). For different values of the cutoff $M_\text{lim}$ the
boosted signal scales like~$\propto M_\text{lim}^{-0.226}$, whereas its
angular profile remains unchanged.

\subsection{Data Analysis}
\label{sec:analysis}
The gamma-ray events entering our analysis are selected from the
\texttt{P7SOURCE\_V6} event class of the Fermi LAT data measured between 4 Aug
2008 and 21 Jul 2011.\footnote{The event data as well as the corresponding
information about the instrument response functions \texttt{P7V6} can be
obtained from \url{http://fermi.gsfc.nasa.gov/ssc/data/}. We checked that
using the event class \texttt{P7CLEAN\_6} instead leads to results that are
similar to what is presented in this paper.} From all events recorded by the
Fermi LAT, we select those with energies between 400\,MeV and 100\,GeV and
apply the zenith angle criterion $\theta<100^\circ$ in order to avoid
contamination by the Earth's Albedo.\footnote{These selections are made using
the \textsf{Fermi Science Tools v9r23p1}. For the cuts in \texttt{gtmktime} we
took \texttt{DATA\_QUAL==1} as well as the RIO-based zenith angle cut.} For
each galaxy cluster, we consider photons events in a $10^\circ\times10^\circ$
squared region centered on the cluster position. These events are binned into
a cube of $0.1^\circ\times0.1^\circ$ pixels with 24 logarithmic energy bins.
The lower end of the considered energy range is somewhat larger than what was
used in previous works, \fex Refs.~\cite{Abdo:2010ex, Ackermann:2010rg}: Below
energies of $400\MeV$, the point spread function (PSF) of the LAT becomes of
the size of our considered target regions~\cite{Fermi:performance}, whereas in
the considered dark matter scenarios no relevant gamma-ray fluxes below
$\sim400\MeV$ are expected; this motivates our choice.

For the diffuse background fluxes we take the isotropic emission and the
galactic foreground model templates currently advocated by the Fermi LAT
collaboration for point source analysis (\texttt{iso\_p7v6source} and
\texttt{gal\_2yearp7v6\_v0}). The galactic foreground model contains several
dedicated spatial templates to model diffuse emission that is not accounted
for by the GALPROP code (\fex for Loop~I and the Galactic
Lobes~\cite{Su:2010qj}).\footnote{See
\url{http://fermi.gsfc.nasa.gov/ssc/data/access/lat/Model_details/Pass7_galactic.html}}
These spatial templates exhibit sharp edges, and we exclude clusters that
coincide with these edges (Centaurus and M49) from our analysis in order to
avoid a bias of our results. On top of the diffuse templates, we add the point
sources from the second Fermi LAT catalog 2FGL~\cite{Collaboration:2011bm}. We
include all point sources within a radius of $12^\circ$ around the cluster
centers. Some of the sources lie outside of our target regions but might still
contribute due to the large point-spread function (PSF) of the Fermi LAT at
low energies.  We furthermore (and conservatively for the purpose of deriving
limits) assume that gamma-ray emission due to \fex shock accelerated cosmic
rays inside the cluster is absent and attribute possible observed fluxes
entirely to dark matter.

We use the profile likelihood method to fit the data and derive
limits~\cite{Rolke:2004mj, Conrad:2007zza}.  The corresponding likelihood
function $\mathcal{L}$ is---for an individual clusters $j$---given by
$\mathcal{L}_j({\boldsymbol c}^j|{\boldsymbol\mu}^j) = \Pi_i
P(c_i^{j}|\mu_i^j)$, where $P(c|\mu)$ denotes the Poisson probability to
observe $c$ events when $\mu$ are expected. The number of expected counts
$\mu_i^j=\mu_i^j({\boldsymbol \alpha})$ that is predicted for an
energy/spatial bin~$i$ is a function of the model parameters $\boldsymbol
\alpha$. These numbers are in principle obtained by a convolution of the above
model fluxes with the instrument response function of the Fermi LAT. In this
work, the convolution with the PSF is done using \texttt{gtsrcmaps} from the
\textsf{Fermi Science Tools}. Like in most of the existing analyses of Fermi
LAT data (for exceptions see Refs.~\cite{Scott:2009jn, Abdo:2010nc,
Vertongen:2011mu}) we will neglect the small but finite energy dispersion of
the LAT, which would not significantly affect the broad energy spectra that we
are considering. Finally, the best-fit model parameters for cluster $j$ are
obtained by maximizing $\mathcal{L}_j({\boldsymbol c}^j| {\boldsymbol
\mu}^j({\boldsymbol\alpha}))$ with respect to the model parameters
$\boldsymbol\alpha$.\medskip

\begin{figure}[t]
  \begin{center}
    \includegraphics[width=0.4\linewidth]{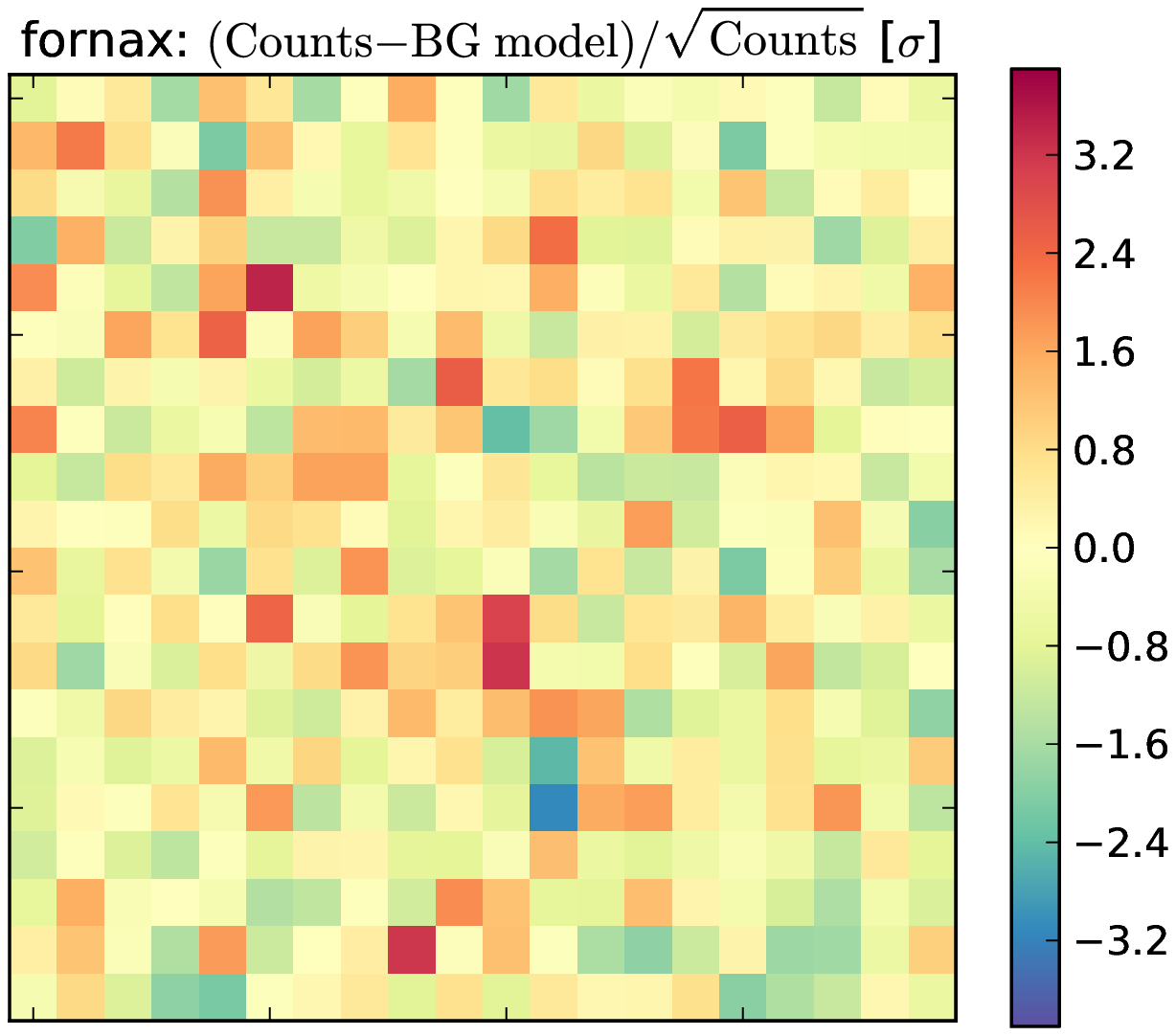}
    \hspace{0.5cm}
    \includegraphics[width=0.4\linewidth]{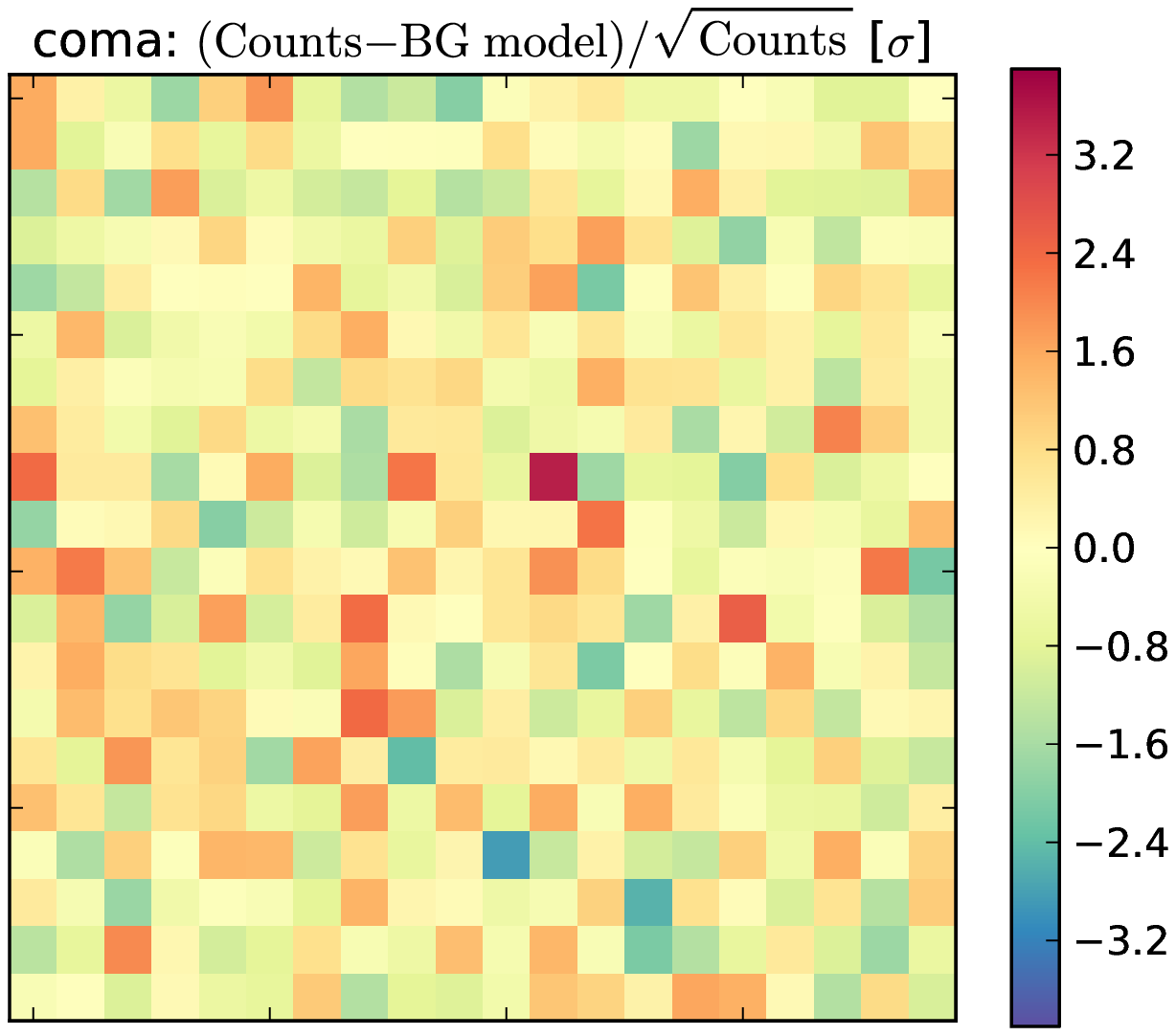}
    \vspace{-0.5cm}
  \end{center}
  \caption{Residual maps after subtraction of our best-fit background models,
  in units of $1\sigma$ standard deviations, for the case of the Fornax and
  Coma cluster. The maps span a $10^\circ\times10^\circ$ region and are
  centered on the cluster position, pixels are resampled to
  $0.5^\circ\times0.5^\circ$, counts are summed over the full energy range
  $400\MeV$--$100\GeV$. The count number per resampled pixel ranges between 11
  and 273 with an average of 28.0.}
  \label{fig:residual}
\end{figure}

\begin{figure}[t]
  \begin{center}
    \includegraphics[width=0.51\linewidth]{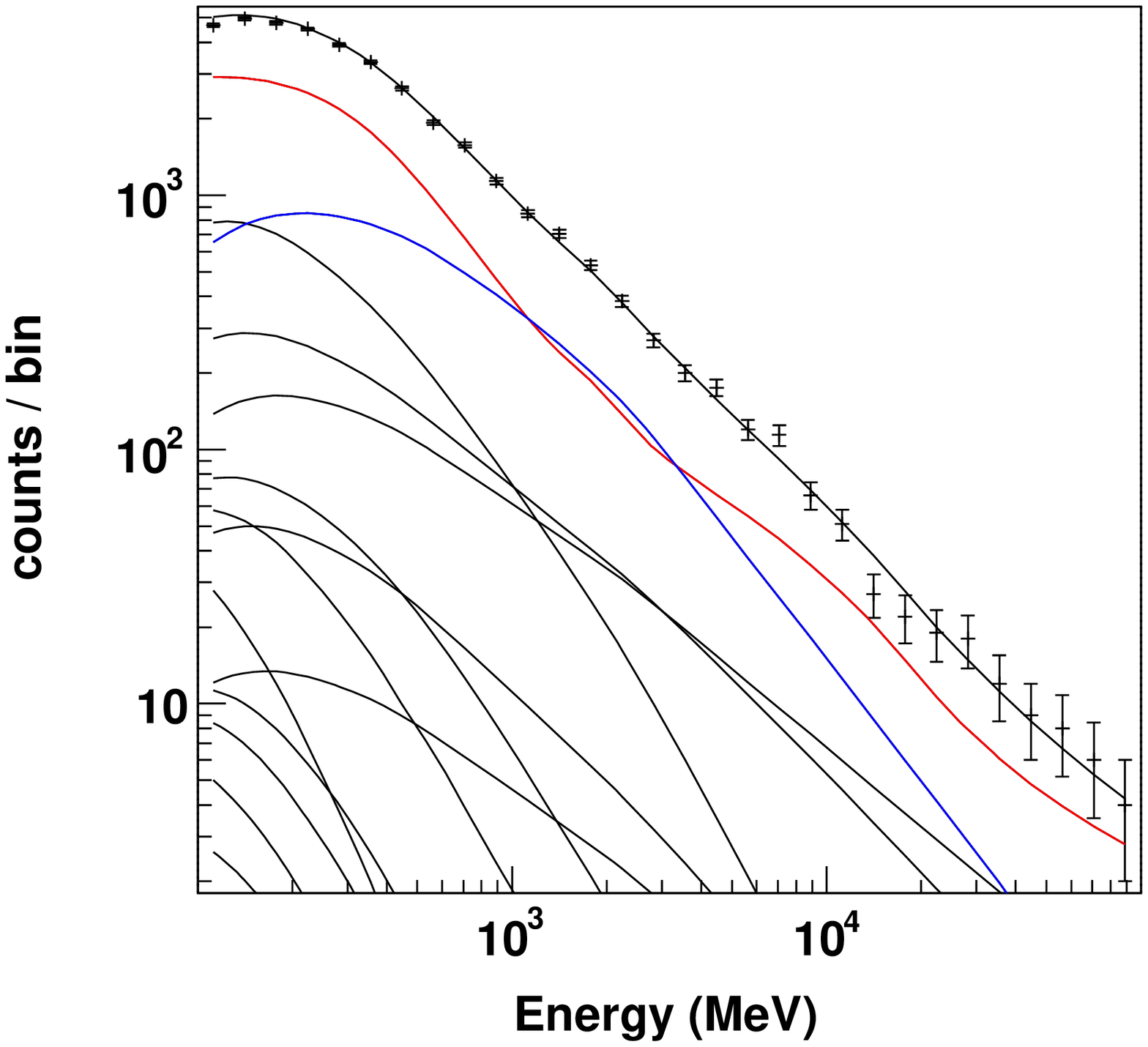}
    \hspace{-0.6cm}
    \includegraphics[width=0.51\linewidth]{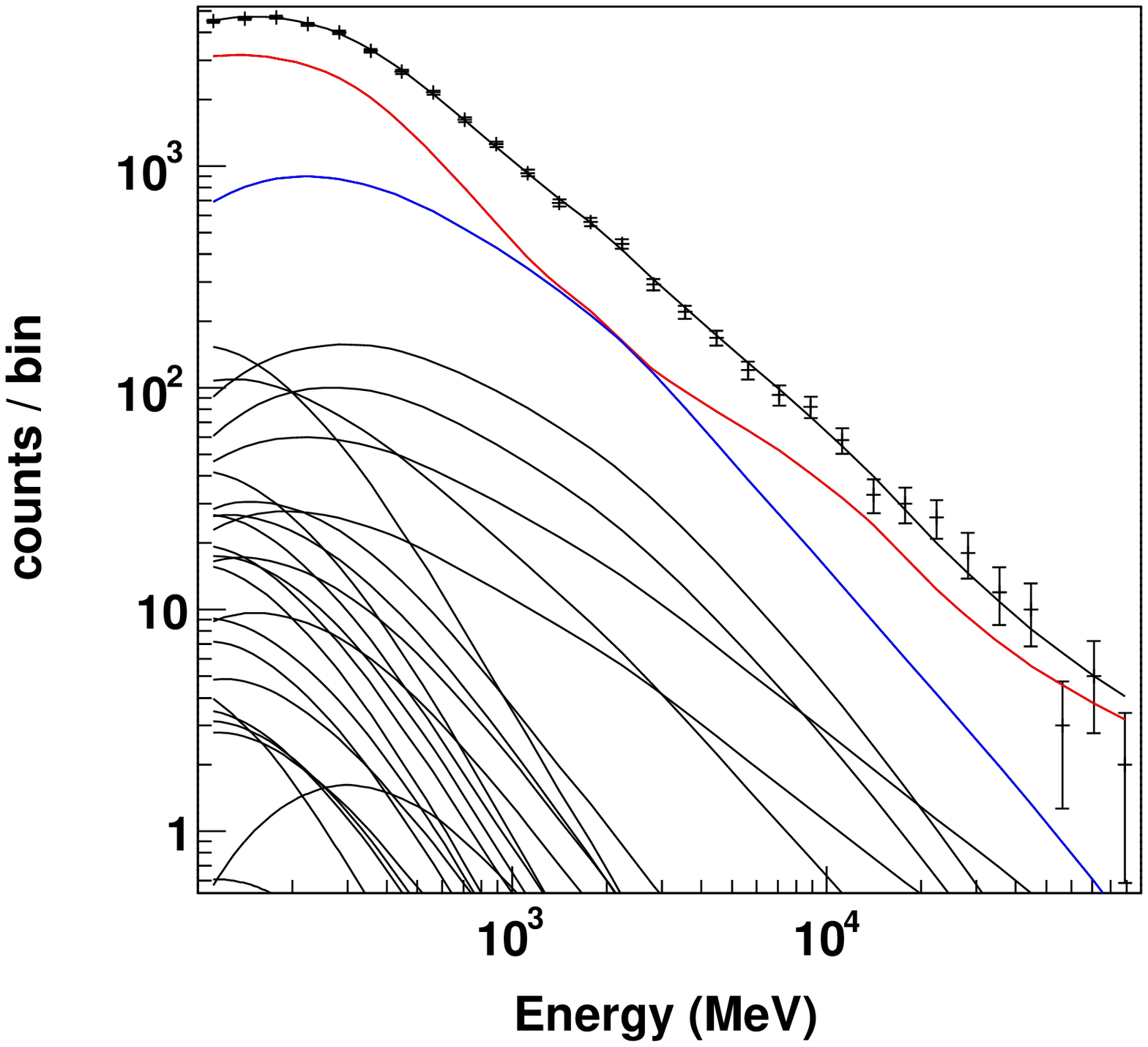}
    \vspace{-1.0cm}
  \end{center}
  \caption{Energy spectra of different background components compared with
  data in the energy range $100\MeV$--$100\GeV$, for Fornax (left panel) and
  Coma cluster (right panel). Blue and red lines correspond to our two diffuse
  templates \texttt{gal\_2yearp7v6\_v02} and \texttt{iso\_p7v6source},
  respectively, the lower black lines show the contribution from different
  point sources of the 2FGL. When performing the fits, we only use data down
  to $400\MeV$, but the background models continuously connect also to data at
  lower energies, as shown in the plot.}
  \label{fig:bgspec}
\end{figure}

As a first step, we fit the data extracted from our eight target regions with
the background model only. The free parameters in the fit are  the
normalizations of the two diffuse background templates, as well as the
normalization and spectral index of all point sources inside a $5^\circ$
radius around the cluster position or with $TS$-values larger than 9. In
Fig.~\ref{fig:residual} we show for two exemplary clusters the residual maps
that we obtain after subtracting our best-fit background models from the data,
integrated over all energies; the corresponding energy spectra of the
individual background components are shown in Fig.~\ref{fig:bgspec}.  For
other clusters, we obtain similar results. The figures indicate that the
adopted background models are sufficient to model the observations. 

\medskip


We then include the potential dark matter signals in the fits. The individual
cluster signals are modeled as extended sources; their surface densities
follow from Eqs.~\eqref{eqn:fluxDDM} and~\eqref{eqn:fluxADM} and are plotted
in Fig.~\ref{fig:profile}. We neglect the smooth Galactic and extragalactic
components (\cp Fig.~\ref{fig:profile}, dotted line) of the dark matter signal
and assume that they are already accounted for by our two diffuse
templates.\footnote{This is a realistic assumption, since these contributions
to the dark matter signal could be easily mistaken as part of the
extragalactic gamma-ray background as determined by the Fermi LAT
collaboration~\cite{Abdo:2010nz}.  Since we already include a template for
this extragalactic flux, a further inclusion would lead to double counting.}
Uncertainties of the cluster masses as given in Ref.~\cite{Chen:2007sz}
translate into uncertainties on the integrated signal from each cluster, and
into uncertainties on its angular shape as parametrized by $\theta_s$; the
resulting errors are indicated in Tab.~\ref{tab:clusters}. We find that
variations in $\theta_s$ have much less impact on our limits than variations
of the integrated signal (less than 10\% for the ranges given in
Tab.~\ref{tab:clusters}); for simplicity we will keep $\theta_s$ at its
central value when performing fits to the data.  

We include uncertainties of the cluster mass as a systematic error into the
profile likelihood method by substituting the likelihood function of an
individual cluster $\mathcal{L}_j$ (with a dark matter signal modeled
according to the central values of Tab.~\ref{tab:clusters}) with a likelihood
function that takes into account the corresponding uncertainties of the
integrated dark matter signal~\cite{collaboration:2011wa}
\begin{equation} 
  \mathcal{L}_j(\alpha_\text{DM})\to 
  \mathcal{\bar L}_j(\alpha_\text{DM})\equiv
  \text{max}_{J^j_{\Delta
  \Omega}} \mathcal{L}_j\left(\alpha_\text{DM}\frac{J^j_{\Delta
  \Omega}}{\bar J^j_{\Delta \Omega}}\right) \mathcal{L}^{\Delta M}_j(J^j_{\Delta
  \Omega})\;.  \label{}
\end{equation} 
Here, $\alpha_\text{DM}$ denotes the normalization of the dark matter signal
(being related to the dark matter lifetime or annihilation cross-section),
$J_{\Delta\Omega}^j$ is the integrated surface density, $\bar
J_{\Delta\Omega}^j$ its central value as given in Tab.~\ref{tab:clusters}, and
$\mathcal{L}^{\Delta M}_j$ is the likelihood function of $J^j_{\Delta\Omega}$
for cluster $j$.  In this work, we approximate $\mathcal{L}^{\Delta M}_j$ by a
log-normal distribution that is defined according to the error bars in
Tab.~\ref{tab:clusters} in order to model the uncertainties of
$J_{\Delta\Omega}^j$. To this end, we fix $\mathcal{L}^{\Delta M}_j$ such that
its cumulative distribution function equals $0.16$ and $0.84$ at the lower and
upper errors given in Tab.~\ref{tab:clusters}, respectively.\footnote{Redoing
the error analysis that was performed in Ref.~\cite{Chen:2007sz}, we found
that a log-normal function describes well the posterior probability
distribution function (pdf) of the cluster masses (as long as the polytropic
index is $\gamma\approx1$).  Since $J^j_{\Delta\Omega}\propto
(M_{500})^\alpha$ with $\alpha\approx1$, the same holds for the posterior pdf
of $J^j_{\Delta\Omega}$. Assuming a flat prior for the cluster masses, this
motivates us to adopt a log-normal function also for the likelihood function
of $J^j_{\Delta\Omega}$.} 

In our signal+background fit, we fix most of the 2FGL sources to their values
from the above background-only fit; exceptions are the sources 2FGL
J1037.5-2820, 2FGL J0334.3-3728 and 2FGL J1505.1+0324, which lie close to the
A1060, Fornax and NGC5813 cluster positions, respectively. This leaves us for
most of the clusters with three free parameters: the signal normalization
$\alpha_\text{DM}$ and the two normalizations of the diffuse backgrounds. We
checked that leaving more 2FGL source parameters free in the fits does not
change our results significantly, but increases the computational time
considerably.  We scan the likelihood function $\mathcal{L}_j^{\Delta M}$ as
function of $\alpha_\text{DM}$ while refitting the remaining free parameters.
Upper limits at the $95\%$ C.L.~($99.7\%$ C.L.) on the dark matter signal can
be derived by increasing the signal until $-2\log\mathcal{L}_j^{\Delta M}$
increases by 2.71 (7.55) from its best-fit value.  The significance of a
signal can be obtained by comparing the likelihood values that are obtained
with and without a dark matter signal.
\medskip

Finally, to combine the statistical power of the different target regions and
to reduce the impact of the cluster mass uncertainties, we performed a
\textit{combined likelihood analysis} of all eight clusters simultaneously.
In this case, the combined likelihood function $\mathcal{L}_\text{comb}$ is
defined as the product of the individual likelihood functions,
$\mathcal{L}_\text{comb}=\Pi_j \mathcal{\bar L}_j$, where $j$ runs over the
different galaxy clusters.  The only parameter that is bound to be identical
for all targets is the dark matter lifetime or annihilation
cross-section.\footnote{Note that the angular distance between the targets
A1060 and S636 is only $8^\circ$ and hence their target regions, but not the
signal regions, overlap to a certain degree. We checked that when profiling
over the signal normalization up to the $2\sigma$ limits (in order to obtain
$\mathcal{\bar L}_j$) the background normalizations are only affected at the
$<1\%$ level, hence the limits on A1060 and S636 remain practically
statistically decoupled and the combined likelihood analysis is applicable.}
\medskip

Note that we use our own software to profile over the combined likelihood
function in presence of cluster mass uncertainties. These scanning routines
were implemented on top of the \textsf{Fermi Science Tools} (and are
independent of the routines used in the combined dwarf analysis of
Ref.~\cite{collaboration:2011wa}).

\section{Limits from the EGBG}
\label{sec:EGBG}
In case of dark matter decay, an important contribution to the gamma-ray
signal always comes from our own Galaxy.  Assuming an NFW profile
($r_s=20\kpc$ and $\rho_\odot=0.4\GeV/\cm^3$), we obtain
$J^\text{dec.}_{|b|>10^\circ}=2.1\cdot10^{22}\GeV\cm^{-2}\sr^{-1}$ when
averaging over the whole sky excluding the Galactic disk, and
$J^\text{dec.}_{\ell=180^\circ}=1.1\cdot10^{22}\GeV\cm^{-2}\sr^{-1}$ at the
Galactic anti-center (the maximal isotropic component of the Galactic flux).
In Fig.~\ref{fig:profile} we compare the angular profiles of the cluster decay
signal with the contribution from our Galaxy in anti-center direction. As
evident from this plot, the Galactic component dominates the signal already at
a distance above $\sim0.5^\circ$ from the cluster center.

For comparison with our galaxy cluster limits, we will derive additional
limits on decaying dark matter by requiring that the isotropic component of
the Galactic signal plus the spatially averaged extragalactic signal does not
overshoot the extragalactic gamma-ray background (EGBG) as derived by the
Fermi LAT collaboration~\cite{Abdo:2010nz} (see \fex
Refs.~\cite{Bertone:2007aw, Ishiwata:2009dk, Papucci:2009gd,Cirelli:2009dv,
Pohl:2009qt, Zhang:2009ut}). In the calculation of the prompt signal
component, we fully take into account the Galactic (in anti-center direction)
and the red-shifted extragalactic signal flux, and we employ for completeness
the inter-galactic background light model of Ref.~\cite{Stecker:2005qs} for
modeling absorption effects. However, our limits do not depend much on the
adopted background light model,\footnote{The adopted background light model
appears to be in conflict with recent Fermi LAT observations, see
Ref.~\cite{Abdo:2010kz}.} since they are dominated by the Galactic signal in
most cases (for details of the calculation see Ref.~\cite{Ibarra:2009nw}).
When calculating the ICS component, however, we conservatively only include
the extragalactic part, coming from electrons/positrons from dark matter decay
that scatters on the CMB. The calculation of the ICS emission inside our
Galactic diffusion zone is plagued with uncertainties and a detailed study is
beyond the scope of this paper (see \fex~Ref.~\cite{Zhang:2009ut} for a
thorough discussion). When quoting limits, we will require that in none of the
energy bins considered in Ref.~\cite{Abdo:2010nz} the dark matter signal
integrated over these bins exceeds the measured flux by more than $2\sigma$.
Such limits can be further improved by performing spectral
fits~\cite{Abdo:2010dk} or subtracting known astrophysical contributions to
the extragalactic gamma-ray background~\cite{Abazajian:2010zb, Calore:2011bt}.

\section{Results}
\label{sec:results}
\begin{figure}[h]
  \begin{center}
    \includegraphics[width=\linewidth]{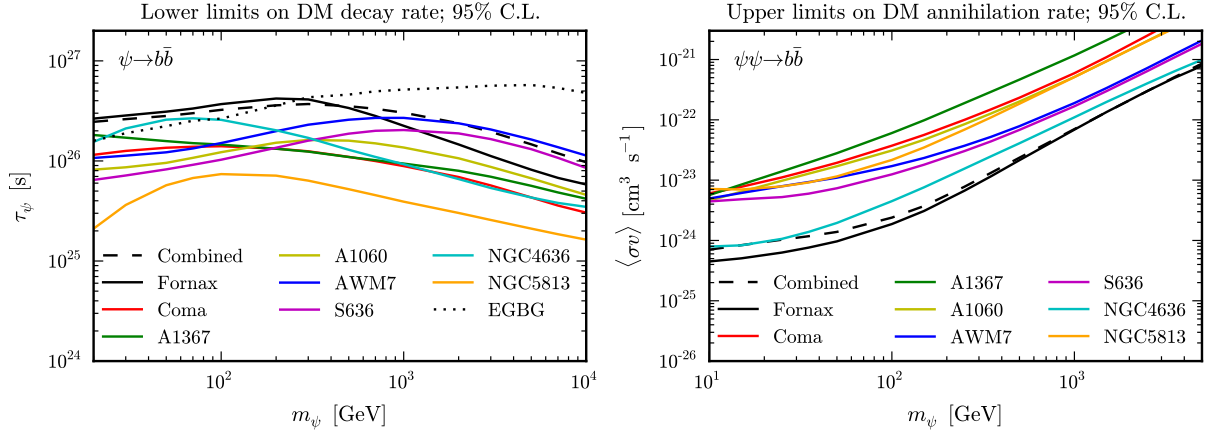}
  \end{center}
  \caption{Left panel: lower limits on dark matter lifetime for decay into
  $b\bar{b}$ final states, as function of the dark matter mass. Solid lines
  show individual cluster limits, the dashed line the limits from the combined
  likelihood analysis. The dotted line shows for comparison the limit that can
  be derived from the EGBG.  Right panel: like left panel, but upper limits on
  annihilation cross-section. Contributions to the signal from dark matter
  substructures are neglected. Note that the combined limits are sometimes
  weaker than the strongest individual limits; this is due to a weak
  preference of a non-zero signal for some of the clusters, see discussion in
  text.}
  \label{fig:limitsBBAR}
\end{figure}

\begin{figure}[h]
  \begin{center}
    \includegraphics[width=\linewidth]{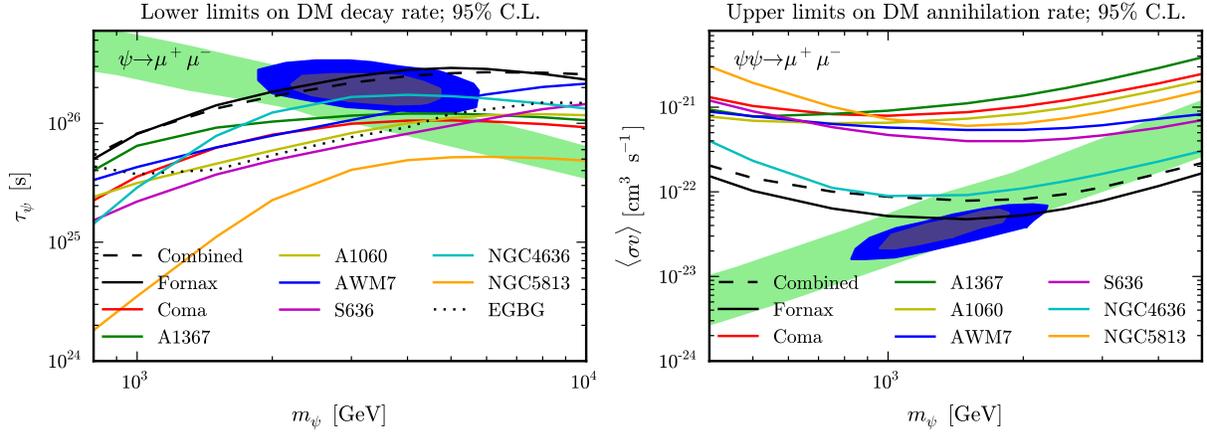}
  \end{center}
  \caption{Like Fig.~\ref{fig:limitsBBAR}, but for $\mu^-\mu^+$ final states.
  The blue region indicates the parameter region where a good fit to the
  PAMELA/Fermi LAT and H.E.S.S.~electron/positron data is
  achieved~\cite{Meade:2009iu}, the green region corresponds to a fit to the
  PAMELA data only.}
  \label{fig:limitsMUMU}
\end{figure}

\begin{figure}[h]
  \begin{center}
    \includegraphics[width=\linewidth]{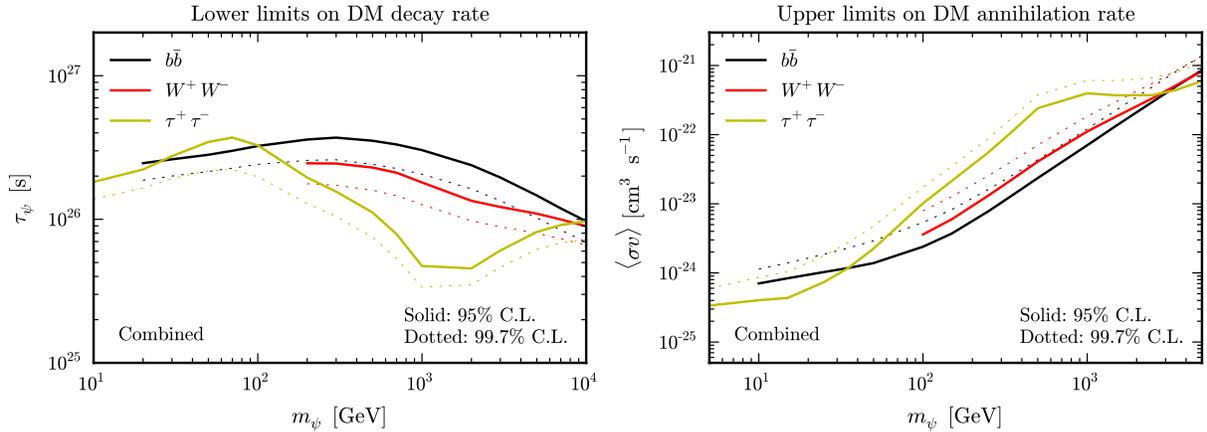}
  \end{center}
  \caption{Left panel: lower limits on dark matter decay rate from the
  combined likelihood analysis, for different final states, as function of the
  dark matter mass. We show limits at $95\%$~C.L.~(solid) and $99.7\%$
  C.L.~(dotted) for comparison. Right panel: corresponding upper limits on
  annihilation rate.  Contributions to the signal from dark matter
  substructures are neglected in this plot.}
  \label{fig:limitsSTACKED}
\end{figure}

\begin{figure}[h]
  \begin{center}
    \includegraphics[width=\linewidth]{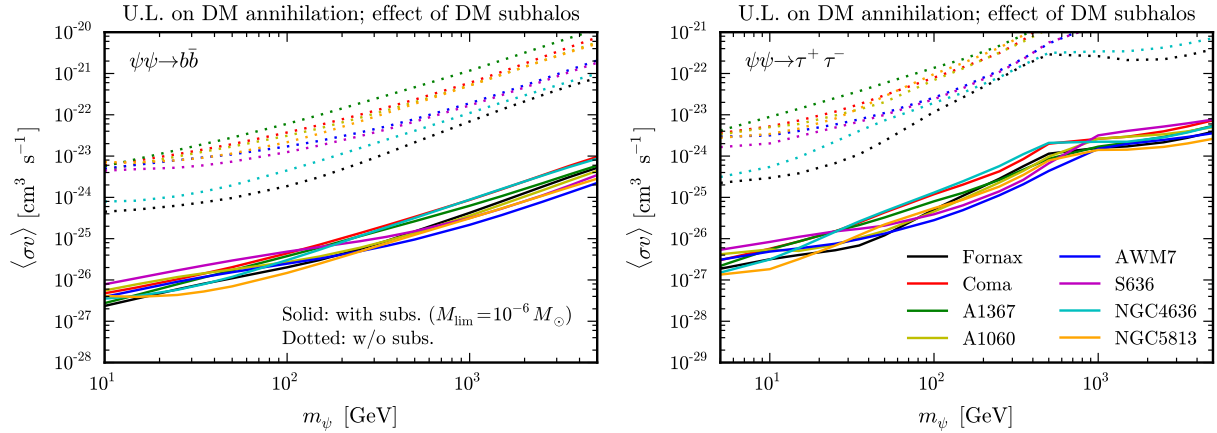}
  \end{center}
  \caption{Left panel: individual upper limits on annihilation rate into
  $b\bar{b}$, including a signal boost from dark matter substructures as
  discussed in Section~\ref{sec:boosts} (solid lines).  For comparison, the
  dotted lines show the limits on the unboosted signal, \cp
  Fig.~\ref{fig:limitsBBAR}. Right panel: the same, but for annihilation into
  $\tau^+\tau^-$. }
  \label{fig:limitsBOOSTED}
\end{figure}

In none of the galaxy clusters a gamma-ray emission was found at the $3\sigma$
level, neither when searching for decay nor for annihilation signals with or
without substructure contributions.\footnote{The best signal candidate comes
from A1367 (annihilation into $\tau^+\tau^-$, $m_\text{DM}=10\GeV$) with a
trial-corrected significance of $2.7\sigma$.}  We derived $95\%$ C.L.~limits
on the dark matter lifetime and annihilation cross-section, from individual
clusters as well as in a combined likelihood analysis; our results are shown
in Figs.~\ref{fig:limitsBBAR}, \ref{fig:limitsMUMU}, \ref{fig:limitsSTACKED}
and \ref{fig:limitsBOOSTED}, assuming $100\%$ branching ratios into
$b\bar{b}$, $\mu^+\mu^-$, $W^+W^-$ or $\tau^+\tau^-$ final states.

\paragraph{Main Results.} For decay or annihilation into $b\bar{b}$, as
relevant for MSSM neutralino DM, the strongest individual limits come from the
Fornax cluster in most cases, as shown in Fig.~\ref{fig:limitsBBAR} (the
impact of dark matter substructures on the annihilation limits is discussed
below).  Depending on the dark matter mass, lifetimes up to $4\cdot10^{26}\s$
and annihilation cross-sections down to $5\cdot10^{-25}\cm^3\s^{-1}$ can be
constrained.  Further strong limits come from AWM7, S636 and NGC4636.  Our
limits on the dark matter lifetime are somewhat weaker than previous
results~\cite{Dugger:2010ys}; the difference can be mainly attributed to the
fact that we modeled the cluster emission as an extended signal rather than as
a point-like source, as we will discuss below. 

\begin{figure}[t]
  \begin{center}
    \includegraphics[width=\linewidth]{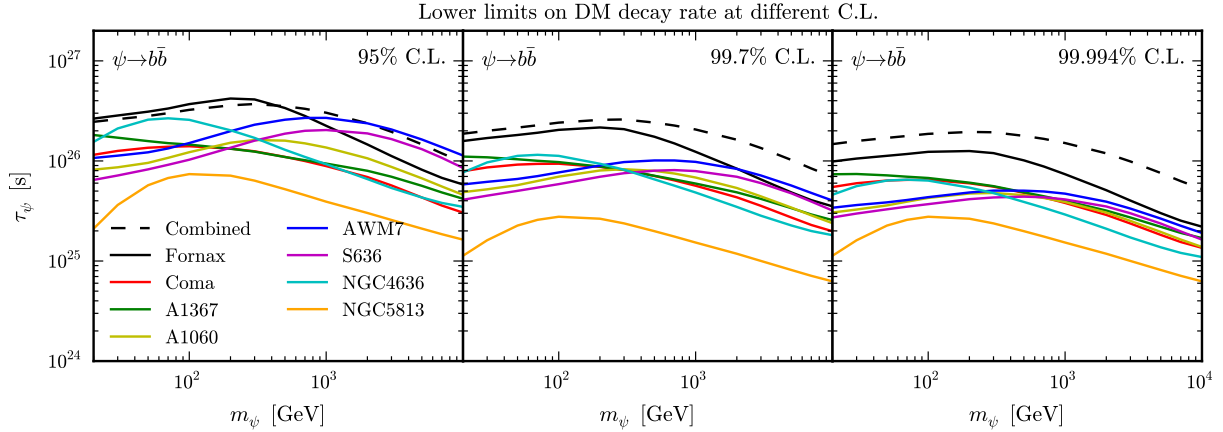}
  \end{center}
  \caption{Like left panel of Fig.~\ref{fig:limitsBBAR}, but at different C.L.
  Note that the combined limit is less dependent on the C.L.~than the
  individual limits, since uncertainties in the cluster masses tend to average
  out if limits are combined.}
  \label{fig:limitsSIGMA}
\end{figure}

The limits obtained from our combined likelihood analysis are shown as dashed
black lines in Fig.~\ref{fig:limitsBBAR}: They are often slightly weaker than
the strongest individual limits. This is due to a weak preference for a
non-zero signal in some of the clusters (like in A1367). In any case the
combined likelihood limit is more robust with respect to uncertainties of the
cluster masses, the background modeling and statistical fluctuations in the
data than the individual limits (see below discussion and
Fig.~\ref{fig:limitsSIGMA}).

For comparison, the dotted line in the left panel of Fig.~\ref{fig:limitsBBAR}
shows the lifetime limits that we obtain from conservatively requiring that
the isotropic dark matter signal does not overshoot the EGBG as determined by
Fermi LAT~\cite{Abdo:2010nz}, see Section~\ref{sec:EGBG}. The EGBG limit
clearly dominates the cluster limits for large dark matter masses, whereas the
cluster limits are competitive for masses below a few hundred GeV. In any
case, since the systematics related to background subtractions are different
for EGBG and cluster limits, the limits should be considered as being
complementary.\medskip

Limits on decay or annihilation into $\mu^-\mu^+$ final states are shown in
Fig.~\ref{fig:limitsMUMU}. This channel is relevant for leptophilic
models~\cite{Fox:2008kb, Kyae:2009jt, Ibarra:2009bm, Bi:2009uj,
Spolyar:2009kx, Cohen:2009fz, Chun:2009zx}, that aim to explain the
PAMELA/Fermi $e^\pm$ anomalies, the corresponding best fit regions being shown
in green (PAMELA only) and blue (PAMELA + Fermi +
H.E.S.S.)~\cite{Meade:2009iu}.\footnote{Recently released Fermi LAT
results~\cite{FermiLAT:2011ab} indicate that the positron fraction continues
to rise up to energies of $200\GeV$, which will presumably shift the prefered
DM mass range to somewhat higher values.} In the presented dark matter mass
range, the dark matter signal is dominated by ICS radiation of the produced
electrons and positrons on the CMB; the prompt final-state-radiation can be
neglected.  Our dark matter lifetime limits reach up to $3\cdot10^{26}\s$ for
individual clusters as well as in the combined likelihood analysis, with the
strongest limit coming from Fornax. In the case of dark matter annihilation
limits down to $6\times10^{-23}\cm^3\s^{-1}$ are obtained. The parameter space
favored by PAMELA/Fermi is constrained but not excluded in case of dark matter
decay, and remains practically unconstrained in case of dark matter
annihilation. 

The dotted line in the left panel of Fig.~\ref{fig:limitsMUMU} shows again the
limit obtained from the EGBG. In the case of decay into $\mu^+\mu^-$, the
cluster lifetime limits actually dominate over our conservative EGBG limit at
all considered dark matter masses. This is due to the fact that we neglected
the Galactic ICS emission when calculating the EGBG limit. Such a calculation
would require a treatment of cosmic-ray propagation in our Galaxy, which has
its own specific uncertainties and is beyond the scope of this paper (see \fex
Ref.~\cite{Zhang:2009ut}). On the other hand, as discussed above, propagation
effects in galaxy clusters can be neglected at angular scales relevant for LAT
observations, making the cluster limits on $\mu^+\mu^-$ final states
practically independent of propagation model uncertainties.\medskip

In Fig.~\ref{fig:limitsSTACKED}, we finally present a summary of our combined
likelihood limits on dark matter decay and annihilation into different final
states, $b\bar{b}$, $W^+W^-$ and $\tau^+\tau^-$. For comparison, we plot the
$95\%$ C.L.~ as well as the $99.7\%$ C.L.~limits. Limits on $W^+W^-$ are up to
mass-independent rescaling similar to the limits on $b\bar{b}$; the limits on
$\tau^+\tau^-$ clearly indicate that at dark matter masses of $\sim1\TeV$ the
ICS part of the dark matter signal starts to dominate inside the considered
gamma-ray energy range.

\paragraph{Dark Matter Substructures.} 
In the above limits we neglected contributions from dark matter substructures
to the dark matter annihilation signal, firstly to obtain very conservative
limits and secondly for the sake of comparison with previous work.  However,
dark matter substructures are a prediction of cold dark matter scenarios and
expected to boost the annihilation signal considerably~\cite{Springel:2008by,
Kuhlen:2008qj, Pinzke:2011ek, Gao:2011rf, SanchezConde:2011ap}. To study their
possible impact on our limits, we follow the prescription presented in
Ref.~\cite{Pinzke:2011ek}, which builds on results from the Aquarius
project~\cite{Springel:2008cc, Springel:2008by} and leaves the free streaming
mass scale $M_\text{lim}$ as the only free parameter (see
Sec.~\ref{sec:boosts} for a discussion). The resulting signal profiles are
plotted in the right panel of Fig.~\ref{fig:profile} by the dotted lines,
where we adopted a free streaming mass of $M_\text{lim}=10^{-6}M_\odot$. As
can be seen from this plot, the boosted signal profiles extend to much larger
radii than the profiles from the smooth dark matter halo alone. The
corresponding boost factors of the overall signal are of the order of $10^3$,
consistent with what is found in Refs.~\cite{Pinzke:2011ek, Gao:2011rf}. We
note, however, that in the recent literature also smaller boost factors for
galaxy clusters were discussed~\cite{Abdo:2010ex, SanchezConde:2011ap} (see
also discussion above), and our adopted boosted fluxes should be considered as
being optimistic but not unrealistic.

No evidence for an extended annihilation signal due to dark matter
substructures was found. In Fig.~\ref{fig:limitsBOOSTED} we show the
corresponding $95\%$ C.L.~limits on the annihilation cross-section compared to
the limits obtained without dark matter substructures taken into account (for
simplicity we neglect uncertainties in the overall cluster mass in case of the
boosted signal). As expected, we find that the limits are improved by a factor
of a few hundred; in the case of dark matter annihilating into $b\bar{b}$ with a
thermal cross-section of $3\times10^{-26}\cm^3\s^{-1}$, dark matter masses
below $\approx150\GeV$ can be excluded. For different values of $M_\text{lim}$
the limits would approximately scale like $\propto M_\text{lim}^{0.226}$.

\begin{figure}[t]
  \begin{center}
    \includegraphics[width=\linewidth]{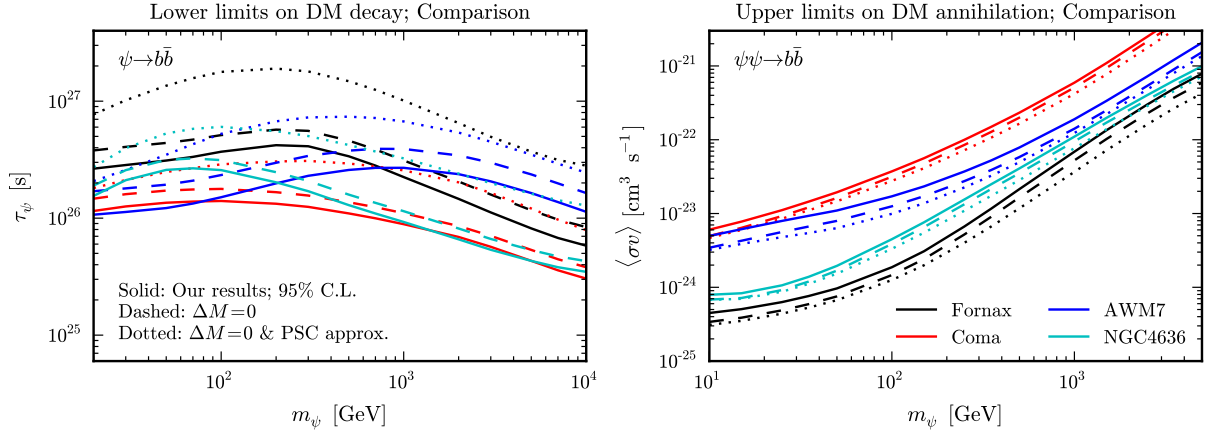}
  \end{center}
  \caption{Like Fig.~\ref{fig:limitsBBAR}, but comparison of some of our
  results (solid lines, \cp Fig.~\ref{fig:limitsBBAR}) with the limits we
  would obtain when neglecting cluster mass uncertainties (dashed lines), and
  when furthermore approximating the cluster emission by a point-like source
  (dotted lines).}
  \label{fig:limitsPSC}
\end{figure}

\paragraph{Discussion.} 
Comparing our results to previously presented limits~\cite{Dugger:2010ys,
Ackermann:2010rg}, we find that despite the increased statistics (at least by
a factor of three) our limits on dark matter decay into $b\bar{b}$ remain
somewhat weaker than the dark matter lifetime limits presented in
Ref.~\cite{Dugger:2010ys}. This is due to the point-source approximation of
the dark matter signal which was underlying the analysis of
Ref.~\cite{Dugger:2010ys}, as well as the inclusion of cluster mass
uncertainties in the present work. In the case of dark matter annihilation our
limits improve the results from Ref.~\cite{Ackermann:2010rg} by up to a
factor of two. 

In Fig.~\ref{fig:limitsPSC} we compare our 95\% C.L.~limits on decay or
annihilation into $b\bar{b}$ (solid lines) with the limits we would obtain
when dropping the cluster mass uncertainties (dashed lines), and when
furthermore approximating the dark matter signal as a point-like source
(dotted lines).  In the latter case, the extended signals are integrated over
a region with a radius of $1^\circ$ and modeled as a point source at the
cluster center, following Ref.~\cite{Ackermann:2010rg, Dugger:2010ys}. As can
be seen from Fig.~\ref{fig:limitsPSC}, the non-inclusion of uncertainties in
the cluster masses could strengthen the limits by up to $\approx50\%$ in some
cases, similar for both dark matter decay and annihilation. A point-source
approximation to the signal from decaying dark matter would lead to a further
strengthening of the corresponding limits by a factor of a few in some cases,
as shown in the left panel of Fig.~\ref{fig:limitsPSC} (a similar impact is
expected in case of annihilation signals boosted by dark matter
substructures).  However, in case of an unboosted annihilation signal (right
panel), the point-source approximation to the signal changes the limits only
on the $10\%$--$30\%$ level and hence appears to be justified, in agreement
with what was found in Ref.~\cite{Ackermann:2010rg}.\medskip

The robustness of a limit with respect to the underlying statistics and
uncertainties can be inferred from its dependence on the adopted C.L. In
Fig.~\ref{fig:limitsSIGMA} we show for the case of dark matter decay into
$b\bar{b}$ our individual and combined limits at $95\%$, $99.7\%$ and
$99.994\%$ C.L.~in comparison. As can be seen from these plots, the combined
limit depends considerably less on the adopted C.L.~than the individual
limits. This is due to the fact that uncertainties related to the cluster
mass, as well as statistical fluctuations in the target region, tend to
average out in the combined likelihood analysis. Furthermore, note that
systematic effects in the modeling of the astrophysical backgrounds become in
general less important at higher C.L. 

\medskip

In this paper we adopted galaxy cluster masses that are based on the X-ray
observations from Refs.~\cite{Reiprich:2001zv, Chen:2007sz}, the HIFLUGCS
catalog. The main advantage of this catalog is its completeness. Many other
X-ray measurements of cluster masses exist (see \fex
Refs.~\cite{McLaughlin:1998sb, Schindler:1998me, Paolillo:2001kb}), which are
however often concentrating on only one single cluster at a time, and hence
difficult to use in a combined analysis. Important assumptions that enter the
mass determination via X-ray observations is that the intra-cluster gas is in
hydrostatic equilibrium and spherically distributed; further assumptions
concern the temperature gradient of the gas, which is often approximated to be
zero, as well as the radial distribution of the gas density. The systematical
errors made by these approximations are very difficult to estimate, and it is
likely that a neglect leads to a too optimistic determination of the cluster
mass uncertainties. 

Fortunately, X-ray measurements are not the only way to determine the mass of
a galaxy cluster; other methods include studies of the velocity dispersion of
cluster member galaxies and weak gravitational lensing (see \fex
Refs.~\cite{Murphy:2011yz, Drinkwater:2000dr, Cote:2003er, Okabe:2010ue}). A
comparison of the masses derived from different methods can be used as a rough
estimate for the overall systematic errors in the cluster mass measurements.
In Refs.~\cite{Ackermann:2010rg} and~\cite{Dugger:2010ys}, the dark matter signal
fluxes that follow from different cluster mass measurements (namely for the
M49, Fornax and Coma clusters) were compared, and it was concluded that the
overall uncertainties are roughly of the order of a factor of $\sim2$ and not
systematically biased with respect to the HIFLUGCS X-ray values.  Looking at
Tab.~\ref{tab:clusters}, one can see that this is somewhat larger than what
follows from the mass uncertainties given in the HIFLUGCS catalog; the latter
are derived from uncertainties in X-ray profile fit, the temperature
measurements and the temperature gradient. Hence, we expect that our limits on
the dark matter annihilation cross section and lifetimes are not the most
conservative ones that one could obtain for individual clusters when
exploiting all systematic uncertainties. The advantage of our combined
analysis is that, in absence of a systematic bias of the HIFLUGCS catalog,
these kind of uncertainties are expected to partially average out.

\medskip

In the above calculation of the dark matter induced ICS emission the possible
impact of \textit{intra-cluster magnetic fields} was neglected; this is only
justified if the magnetic fields remain well below the critical value of
$B_\text{CMB}=3.2\mu\text{G}$. However, since Faraday rotation based
measurements in galaxy clusters find magnetic fields of a few $\mu \text{G}$
(see \fex~\cite{Bonafede:2010xg, Ensslin:2010zh, Bonafede:2011hd}), the
validity of this approximation is not guaranteed. For most of our targets in
Tab.~\ref{tab:clusters} the magnetic fields are not precisely know, which
makes it difficult to systematically include their effects. However, in case
of the very massive Coma cluster the magnetic field was studied in
Ref.~\cite{Bonafede:2010xg}, and a model for the magnetic field profile was
presented. Adopting this model we can calculate how our limits change when
synchrotron emission of the electrons and positrons produced in the dark
matter decay or annihilation is included. We find that in case of the Coma
cluster the limits on dark matter annihilation into $\mu^+\mu^-$ as shown in
Fig.~\ref{fig:limitsMUMU} are weakened by a factor of around two, whereas the
effect on decaying dark matter limits is negligible (the same is true for the
extended annihilation signals coming from dark matter substructures). In the
adopted magnetic field model, the field exceeds the critical value
$B_\text{CMB}$ only close to the cluster center at angles
$\theta\lesssim0.17^\circ$, making the impact on point-source like signals
large and on extended signals small. We conclude that intra-cluster magnetic
fields are unlikely to affect our decaying dark matter limits or the limits on
subhalo-boosted annihilation signals.

\section{Consequences for gravitino dark matter}
The gravitino is the spin-3/2 supersymmetric partner of the graviton. If the
lightest superparticle (LSP), it provides a natural dark matter candidate
\cite{Pagels:1981ke}, the mass of which can vary from $m_{3/2} \sim
\textrm{eV}$ to $\sim \textrm{TeV}$ depending on the details of  the
supersymmetry breaking mechanism.  Gravitinos are produced in the early
universe through 2-to-2 thermal scatterings with an abundance which is
proportional to the reheating temperature $T_R$ after inflation
\begin{align}
  \label{eq:gravitinoabundance}
  \Omega_{3/2}^{\textrm{th}}h^2 = C
  \left(\frac{100\,\textrm{GeV}}{m_{3/2}} \right) \bigg(
  \frac{m_{\tilde{g}}}{1\,\textrm{TeV}} \bigg)^2 \left(
  \frac{T_R}{10^{10}\,\textrm{GeV}} \right)\,,
\end{align}
where $m_{3/2}$ and $m_{\tilde{g}}$ are the gravitino and gluino masses
respectively, and $C\simeq 0.5$ to leading order in the gauge couplings
\cite{Bolz:2000fu, Pradler:2006qh, Buchmuller:2008vw}.\footnote{Note that $C$
has ${\cal O}(1)$ uncertainty due to unknown higher order contributions and
nonperturbative effects~\cite{Bolz:2000fu}.} In addition, gravitinos may also
be produced through the gravitational decay of the NLSP. However, for
$\Omega_{\textrm{NLSP}}h^2 \ll 1$ or $m_{\textrm{NLSP}}\gg m_{3/2}$ the latter
contribution is negligible \cite{Pradler:2006hh}. Moreover, inflaton decay may
also contribute to the production mechanism \cite{Nakayama:2010xf}.  In what
follows, thermal leptogenesis is assumed to be responsible for the generation
of the observed baryon asymmetry.  In such a case, high reheating temperatures
are required, and the dominant gravitino production mechanism is the thermal
one.  Particularly, for $T_R \sim 10^{10}$\,GeV \cite{Fukugita:1986hr,
Buchmuller:2004nz} a gravitino abundance of the order of the observed dark
matter relic density $\Omega_{\textrm{DM}} =0.11$ \cite{Komatsu:2010fb} is
achieved for typical supersymmetric parameters, \textit{i.e.} $m_{3/2} \sim
100$\,GeV and $m_{\tilde g} \sim 1$\,TeV.  However, as it is well known, such
high values of the gravitino mass lead to slow NLSP decays and can
dramatically affect the successful predictions of the standard big bang
nucleosynthesis (BBN) scenario \cite{Ellis:1984er, Sarkar:1995dd,
Kawasaki:2004qu, Pospelov:2006sc, Hamaguchi:2007mp, Pospelov:2008ta,
Kawasaki:2008qe}. 

Among the different scenarios proposed to reconcile thermal leptogenesis,
gravitino dark matter and BBN, a mild violation of $R$-parity inducing a rapid
decay of the NLSP before the onset of the BBN is of interest
\cite{Buchmuller:2007ui}. In such a case, the gravitino is not stable anymore,
but still provides a viable dark matter candidate due to the double
suppression of its decay, by the Planck scale as well as by the small
$R$-parity breaking parameter. Interestingly, this opens up the way to look
for traces of gravitino decays in cosmic-ray fluxes, such as anti-matter
\cite{Ishiwata:2008cu, Ibarra:2008qg} and neutrino \cite{Covi:2008jy}.
Additionally to the intense gamma-ray line arising from the $\psi_{3/2}
\rightarrow \gamma \nu$ two-body decay \cite{Takayama:2000uz,
Buchmuller:2007ui, Lola:2007rw, Bertone:2007aw,
Ishiwata:2008cu,Ibarra:2007wg}, the produced  gamma-ray flux typically
features a continuous component generated by the fragmentation of the Higgs
and gauge bosons. 

In what follows, we apply the above analysis to the decaying gravitino
scenario. Contrarily to gamma-ray lines,  galaxy clusters offer more
sensitivity to large gravitino masses, thus rendering the present analysis
supplementary to our previous gamma-ray lines study \cite{Vertongen:2011mu}.
Following the structure of the latter, we first summarize the bilinear
$R$-parity violation supersymmetric framework considered here. We then present
limits on the size of $R$-parity violation and finally discuss the prospect
for seeing long-lived neutralino and stau NLSPs at the LHC.

\subsection{$R$-parity breaking model}
The supersymmetric standard model with explicit bilinear $R$-parity violation
is specified by the superpotential
\begin{align}
  W = W_{\textrm{MSSM}} + \mu_i H_u l_i\,,
\end{align}
as well as by the soft supersymmetry breaking potential
\begin{align}
  {\cal L} = {\cal L}_{\textrm{soft}}^{\textrm{MSSM}} + B_i H_u \tilde
  l_i + m_{id}^2 \tilde l_i^\dagger H_d + \textrm{h.c.}\,,
\end{align}
where $W_{\textrm{MSSM}}$ and ${\cal L}_{\textrm{soft}}^{MSSM}$ are the
$R$-parity conserving MSSM superpotential and scalar Lagrangian, $H_{u/d}$ are
the up/down-type Higgs doublets, $l_i$ the lepton doublets, and $\mu_i$, $B_i$
and $m_{id}^2$ are the $R$-parity violating couplings. Trading the mass mixing
parameters for $R$-parity breaking Yukawa couplings as proposed in
Ref.\,\cite{Bobrovskyi:2010ps}, the gravitino decay is function of a single
dimensionless parameter $\zeta$, which also enters the decay of the NLSPs of
interest (see Ref.\cite{Bobrovskyi:2010ps} for a definition of $\zeta$ in
terms of the bilinear $R$-parity violating couplings $\mu_i$, $B_i$ and
$m_{id}^2$).

\medskip
Two typical sets of boundary conditions for the supersymmetry breaking
parameters of the MSSM at the grand unification (GUT) scale are investigated
in the following, resulting in two different types of NLSPs.  First we
consider equal scalar and gaugino masses
\begin{align}
  \text{(A)}\qquad  m_0 = m_{1/2},\quad a_0=0, \quad \tan \beta =10\,,
  \label{eq:boundaryA} 
\end{align}
for which the bino-like neutralino $\widetilde \chi_1^0$ is the NLSP.  In the
second one, which corresponds to no-scale models or gaugino mediation, 
\begin{align}
  \text{(B)} \qquad m_0 = 0, \quad  m_{1/2}\neq0,\quad a_0=0, \quad \tan
  \beta =10\,,\label{eq:boundaryB} 
\end{align}
the lightest stau $\tilde \tau_1$ is the NLSP. In both cases, $\tan \beta =
10$ has been chosen as a representative value, and the trilinear scalar
coupling $a_0$ has been set to zero for simplicity.  For both sets of boundary
conditions, the universal gaugino mass $m_{1/2}$ remains as the only
independent variable, and the gaugino masses $M_{1,2,3}$ satisfy the following
relations at the electroweak scale 
\begin{align}
  \frac{M_3}{M_1} \simeq 5.9\,,\qquad   \frac{M_2}{M_1} \simeq 1.9\,.
  \label{eq:gauginomasses}
\end{align}
\medskip

Electroweak precision tests (EWPT) yield important lower bounds on the
superparticle mass spectrum~\cite{Buchmuller:2008vw}.  For a neutralino NLSP,
the most stringent constraint comes from the Higgs potential.  The universal
gaugino mass $m_{1/2}$ is required to be high enough in order for the Higgs
mass to fulfills the LEP lower bound $m_h >
114.4$\,GeV~\cite{Nakamura:2010zzi}. This implies the lower limit $\mN \gtrsim
130$\,GeV.\footnote{Note that $\mN\simeq M_1$ with good
accuracy~\cite{Bobrovskyi:2010ps}.} However, allowing negative $a_0$ or scalar
masses much larger than $m_{1/2}$ at the GUT scale would weaken this limit,
and we will take $\mN>100\GeV$ as a lower bound for the neutralino mass
subsequently.  In the stau NLSP case, the lower bound comes from the absence
of pair production of heavy charged particles at LEP and reads
$m_{\tilde\tau_1}>100$\,GeV~\cite{Nakamura:2010zzi}. Rewriting
Eq.~\eqref{eq:gravitinoabundance}
\begin{align}
  m_{\text{NLSP}} \simeq 310\,\text{GeV} \left( \frac{\xi}{0.2} \right)
  \bigg( \frac{m_{3/2}}{100\,\text{GeV}} \bigg)^{1/2} \left(
  \frac{10^9\,\text{GeV}}{T_R} \right)^{1/2}\;,
  \label{eq:NLSPupbound}
\end{align}
where $\xi \equiv m_{\text{NLSP}}/m_{\tilde g}$ is implicitly fixed by the
supersymmetry breaking boundary conditions~\cite{Buchmuller:2008vw}, we get
absolute upper bounds on the NLSP masses requiring the gravitino to be the
LSP.  In the case of the neutralino NLSP, Eq.~\eqref{eq:NLSPupbound} implies
$\mN \lesssim 690$\,GeV for $\xi = 1/5.9$, and is essentially independent of
$m_0$ and $\tan \beta$.  For the stau NLSP, $\tan \beta = 10$ yields $\xi =
1/6.2$, which consequently leads to the more stringent bounds $\mstau \lesssim
615$\,GeV.  Note that there is a strong dependence on $\tan \beta$ in that
case~\cite{Buchmuller:2008vw}, and that $\xi$ decreases with increasing $\tan
\beta$.

For a typical effective neutrino mass $\widetilde m_1 =10^{-3}$\,eV,
successful thermal leptogenesis requires a minimal reheating temperature of
$T_R \sim 10^9$\,GeV~\cite{Buchmuller:2004nz}.  Using
Eq.\,\eqref{eq:gravitinoabundance} together with a lower bound on the gluino
mass $m_{\tilde g} \gtrsim 815$\,GeV~\cite{Aad:2011aa}, this implies a lower
bound for the gravitino mass $m_{3/2} \gtrsim 30$\,GeV.

\subsection{Limits from galaxy clusters}
\begin{figure}[t]
  \begin{center}
    \includegraphics[width=0.49\linewidth]{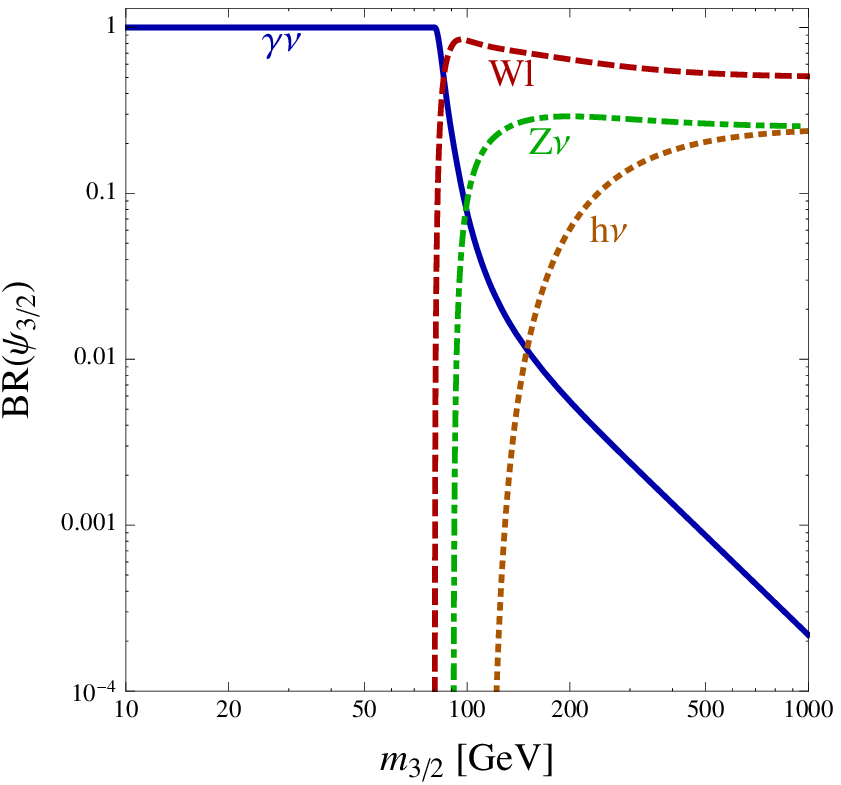}
    \includegraphics[width=0.49\linewidth]{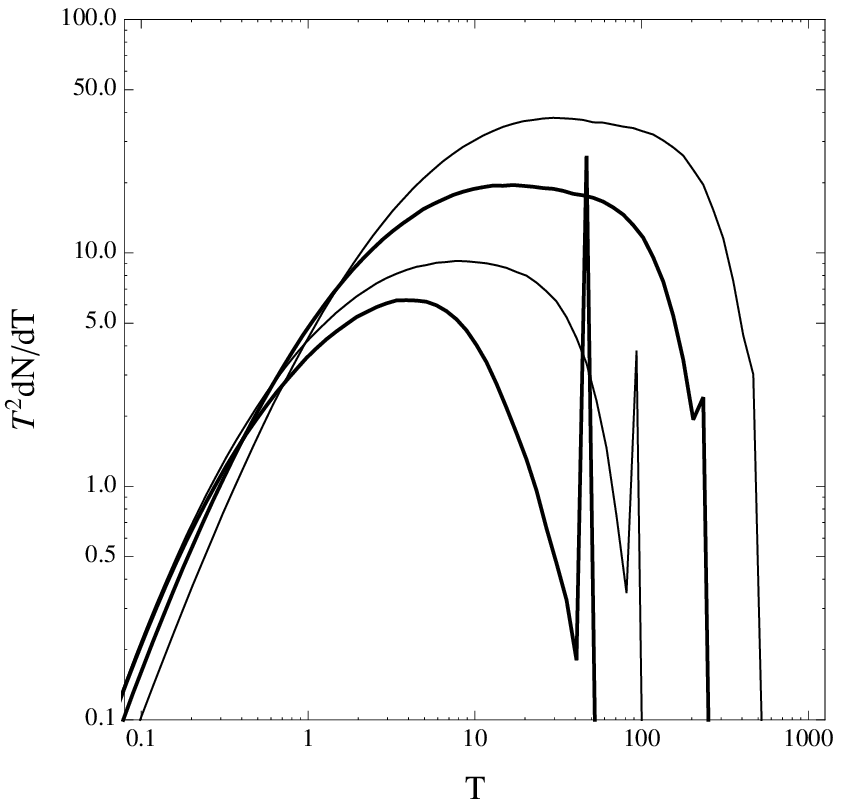}
  \end{center}
  \caption{Left: Two-body decay branching ratios of the gravitino. Right:
  Gamma-ray spectra for $m_{3/2} = 100, 200, 500$ and $1000$\,GeV. We adopt
  here the same set of parameters as in
  Refs.~\cite{Ishiwata:2008cu,Covi:2008jy}. } \label{fig:BR_spectra}
\end{figure}

\begin{figure}[t]
  \begin{center}
    \includegraphics[width=0.7\linewidth]{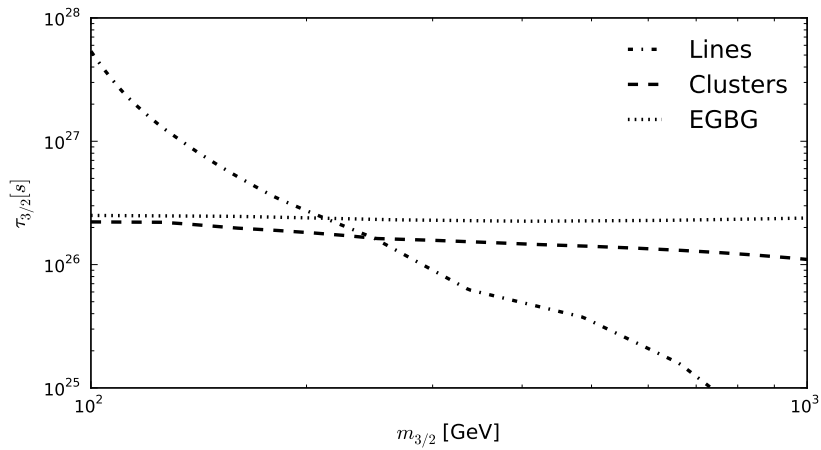}
  \end{center}
  \caption{Lower limits on the gravitino lifetime. The dot-dashed line shows
  the gamma-ray line limits, the dashed line the limits resulting from the
  combined cluster analysis, and the dotted line the EGBG limits.}
  \label{fig:bounds_tau}
\end{figure}

The gamma-ray spectrum produced through gravitino decays features two types of
contributions: First, the $\psi_{3/2} \rightarrow \gamma \nu$ two body decay
produces a gamma-ray line, a channel which is dominant below the $W$
threshold\footnote{Note that three-body decays with intermediate massive gauge
bosons are expected to contribute by more than 10\% below the kinematic
threshold \cite{Choi:2010xn, Choi:2010jt, Diaz:2011pc}.}. Additionally, both
the fragmentation of the Higgs and gauge bosons as in $Z^0 \nu$ and $h^0 \nu$
final states, as well as the final state radiation of the charged leptons
produced in the $W^\pm l^\mp$ final states, generate a continuum spectrum. The
relative strength of these two is fixed by the corresponding branching ratios,
that we present in Fig.~\ref{fig:BR_spectra} together with representative
spectra following Refs.\cite{Ishiwata:2008cu,Covi:2008jy}.\footnote{Note that
the branching ratio into lines is in principle model-dependent.} While the
search for gamma-ray lines of galactic origin through deviations from a power
law background is efficient for contained continuum  contributions
\cite{Vertongen:2011mu}, \textit{i.e.} for $m_{3/2} \lesssim 200-300$\,GeV,
constraints from galaxy clusters observations and the EGBG dominates the
gamma-ray line ones for $m_{3/2} \gtrsim 250$\,GeV, as illustrated in
Fig.~\ref{fig:bounds_tau}. This agreeably makes gamma-ray line searches,
galaxy cluster observations and EGBG studies complementary. As a result the
gravitino lifetime is constrained to be at least $\tau_{3/2} \gtrsim {\cal
O}(10^{26}$\,s) in all the gravitino mass range considered. Subsequently, we
will concentrate on the limits derived by galaxy cluster observations, and
leave a detailed study of implications from the EGBG to future work.

\subsubsection{$R$-parity breaking parameter}
\begin{figure}[t]
  \begin{center}
    \includegraphics[width=0.7\linewidth]{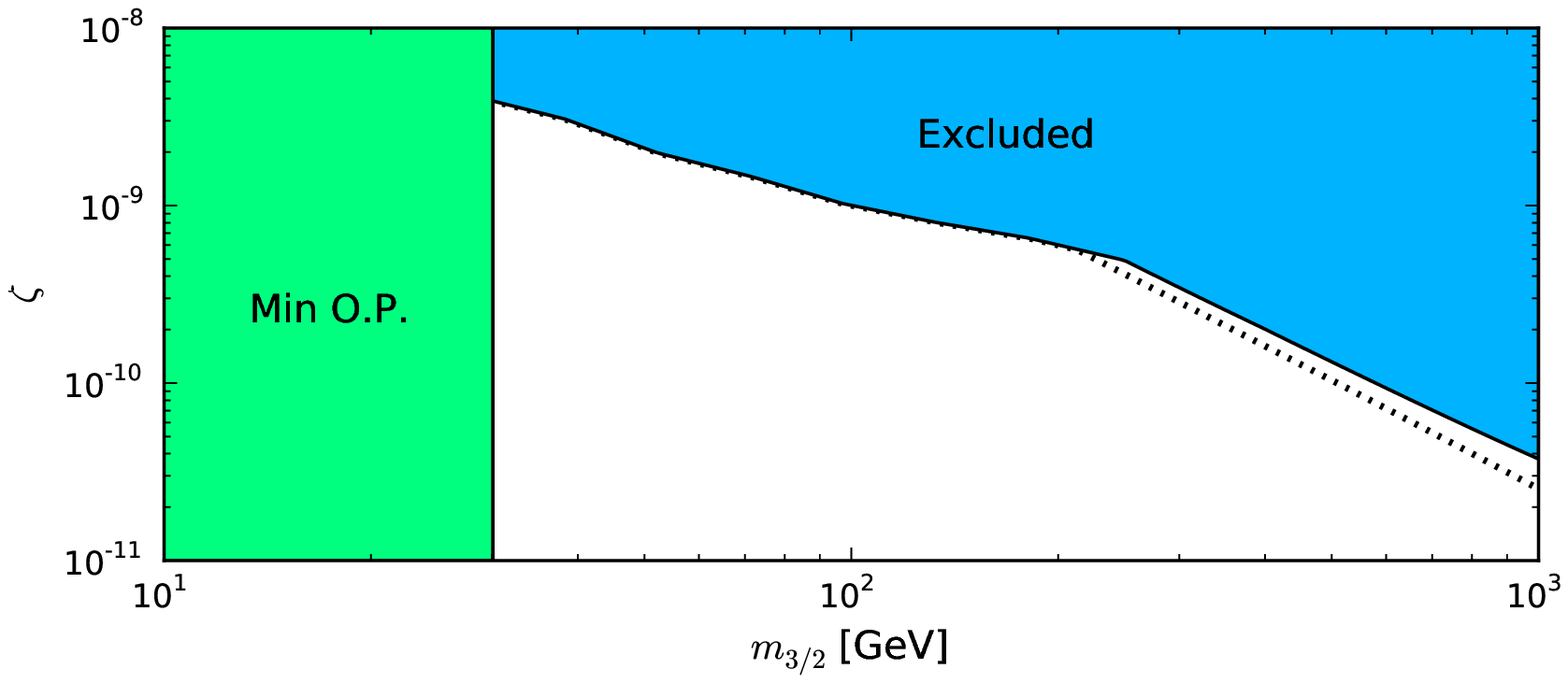}
  \end{center}
  \caption{Upper bounds on the $R$-parity violation parameter $\zeta$, derived
  from the Fermi LAT limits in Fig.~\ref{fig:bounds_tau}. For thermal leptogenesis,
  overproduction (O.P.) of gravitinos already excludes the left green
  region. The dotted line illustrates the limits from EGBG.}
  \label{fig:bounds_zeta}
\end{figure}

The gravitino inverse decay rate into photon/neutrino pairs is given by
\cite{Takayama:2000uz, Bobrovskyi:2010ps}
\begin{align}
  \label{eq:Gdecayrate}
  \Gamma^{-1}_{\psi_{3/2}\to\gamma \nu} = \frac{32 \sqrt{2}}{\alpha \zeta^2}
  \frac{\GF
  \MP^2}{\mG^3} \frac{M_1^2 M_2^2}{(M_2-M_1)^2}\left( 1+ {\cal O}\left( s_{2
  \beta} \,
  \frac{m_Z^2}{\mu^2} \right) \right)\,,
\end{align} 
where $\alpha$ is the electromagnetic fine structure constant, $\MP =
2.4\times 10^{18}$\,GeV the reduced Planck mass, and $\GF = 1.16\times
10^{-5}$\,GeV$^{-2}$ is the Fermi constant. Using the strongest limits on the
total gravitino lifetime illustrated in Fig.~\ref{fig:bounds_tau} together with the branching
ratios presented in Fig.~\ref{fig:BR_spectra}, this expression can be used to
derive conservative upper-limits on the $R$-parity breaking parameter $\zeta$.
To do so, one has to consider for a given gravitino mass the maximally allowed
bino mass which results from the  combination of
Eqs.\,\eqref{eq:gravitinoabundance} and \eqref{eq:gauginomasses} when
considering the lowest reheating temperature allowed in the thermal
leptogenesis scenario, \textit{i.e.} $T_R \sim 10^9$\,GeV. The results are
presented in Fig.~\ref{fig:bounds_zeta}. Note that at high gravitino masses,
the production of anti-protons in Higgs and gauge bosons fragmentation could
further constrain the $\zeta$ parameter.

\subsubsection{Stau NLSP decay length}
In the case of the $\tilde \tau_1$-NLSP, the total decay width of the lightest
mass eigenstate is a mixture of left and right handed partial decays
\begin{align}
  \Gamma_{\tilde \tau_1}(\epsilon) = \sin^2\theta_{\tilde \tau}
  \,\Gamma_{\tilde \tau_L}(\epsilon) 
  + \cos^2\theta_{\tilde \tau} \,\Gamma_{\tilde \tau_R}(\epsilon)\,.
  \label{eq:ctaustau}
\end{align}
where the dimensionless parameter $\epsilon$ is directly related to the
$R$-parity violating Yukawa couplings (see Ref.~\cite{Bobrovskyi:2010ps} for
details).  Since the latter are typically proportional to the ordinary Yukawa
couplings, decays into second and third families dominate.  For definiteness,
we will below assume a flavor structure as described in
Ref.~\cite{Bobrovskyi:2010ps}, into which the chiral state decays are
dominated by the following channels
\begin{subequations}
  \begin{align}
    \tilde \tau_R &\rightarrow \tau_L \nu, \mu_L \nu\,,\\
    \tilde \tau_L & \rightarrow \bar t_R b_L\,.
  \end{align}
\end{subequations}

\begin{figure}[t]
  \begin{center}
    \includegraphics[width=0.80\linewidth]{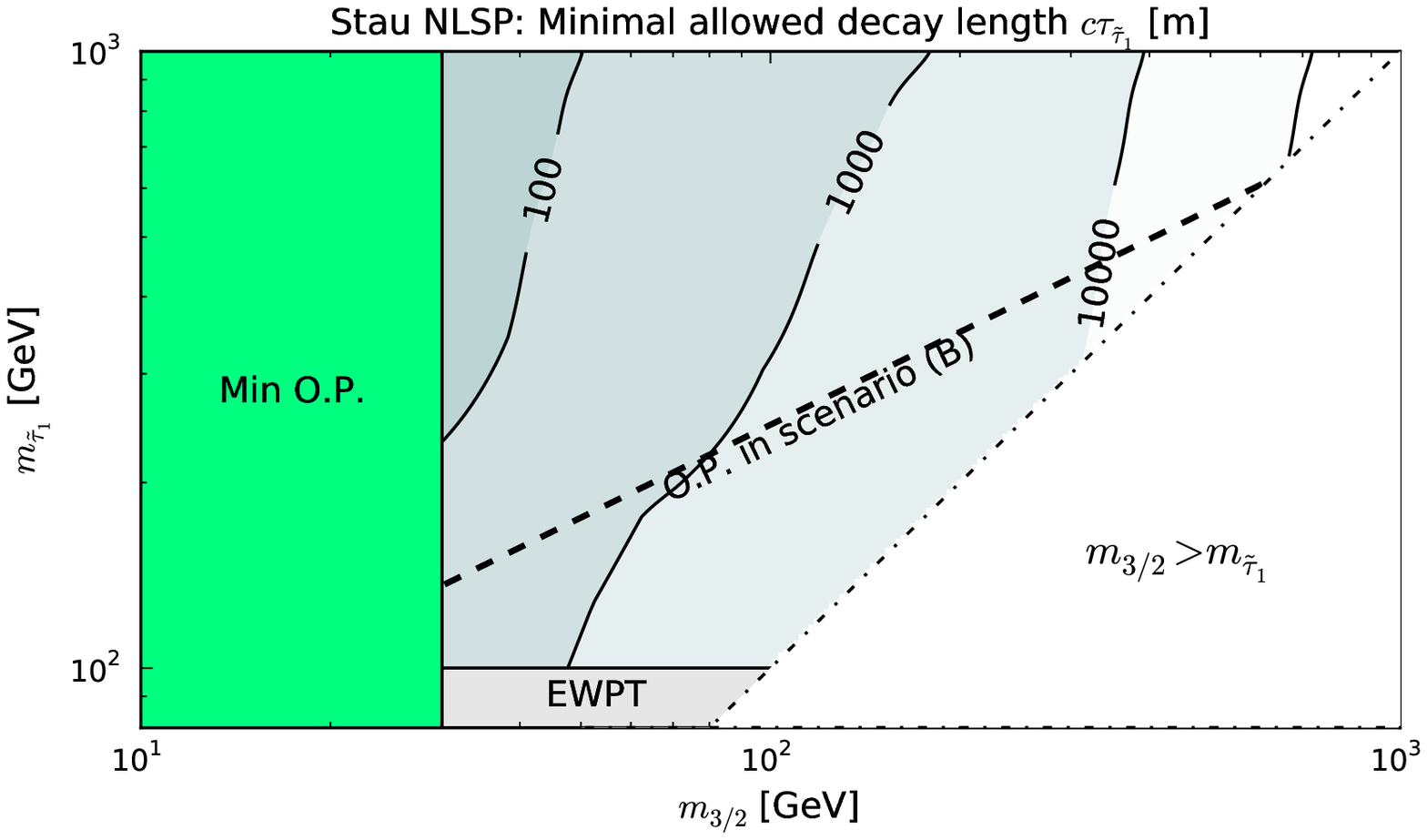}
  \end{center}
  \caption{Contour plot of lower bounds on the stau NLSP decay length coming
  from cluster and gamma-ray line constraints on the gravitino lifetime, as
  function of the stau and gravitino masses, $m_{\tilde\tau_1}$ and $m_{3/2}$
  respectively. The lower gray region is excluded by electroweak precision
  tests (EWPT).  For thermal leptogenesis, overproduction (O.P.) of gravitinos
  excludes at minimum the left green region, a limit which strengthens to the
  black-dashed line when assuming the universal boundary conditions (B),
  \textit{cf.}~Eq.~\eqref{eq:boundaryB}. The lower-right exclusion comes from
  the gravitino LSP requirement.}
  \label{fig:Staudecaylength}
\end{figure}

Assuming $\zeta\simeq\epsilon$,\footnote{$\zeta$ values much smaller than
$\epsilon$ can be achieved through a proper choice of the parameters $\mu_i$,
$B_i$ and $m_{id}^2$.} and using the cluster upper limits on $\zeta$ from
Fig.~\ref{fig:bounds_zeta}, we can derive lower bounds on the stau decay
length.  Our results are shown in Fig.~\ref{fig:Staudecaylength}.  The
parameter space is already constrained by EWPT and overproduction bounds, and
the lower limits on the neutralino decay length vary between 100\,m and
10\,km.  It is interesting that if such particles were to be produced at the
LHC, a sizable amount of their decays could take place in the
detector~\cite{Ishiwata:2008tp,Bobrovskyi:2011vx}.  We obtain a lowest
possible decay length $c\tau_{\tilde \tau_1} \simeq 200$\,m  for $\mG \simeq
30$\,GeV and $\mstau \simeq 130$\,GeV. 

\subsubsection{Neutralino NLSP decay length}
A neutralino NLSP heavier than 100\,GeV dominantly decays into $W^\pm\ell^\mp$
and $Z^0\nu$~\cite{Mukhopadhyaya:1998xj, Chun:1998ub}.  The corresponding
decay width is directly proportional to the $R$-parity breaking parameter
$\zeta$ squared, which also enters the gravitino decay width
Eq.~\eqref{eq:Gdecayrate}.  As a consequence, the two quantities can be
related through~\cite{Bobrovskyi:2010ps}
\begin{align}
  \tau_{\tilde \chi_1^0} = \frac{c_w^2}{2 \sqrt{2}} \frac{(M_2-M_1)^2}{M_2^2}
  \frac{\mG^3}{\GF \MP^2 \mN^3} \,\frac{\Gamma_{\psi_{3/2}\to\gamma\nu}^{-1}}{
  2f(\mN,m_W) + f(\mN,m_Z)}\,,
  \label{eq:ctauneutralino}
\end{align}
where the phase space factor $f$ is defined by
\begin{align}
  f(m_1,m_2) = \left(1- \frac{m_2^2}{m_1^2} \right)^2 \left( 1+
  2\,\frac{m_2^2}{m_1^2}\right)\,.
\end{align}

\begin{figure}[t]
  \begin{center}
    \includegraphics[width=0.80\linewidth]{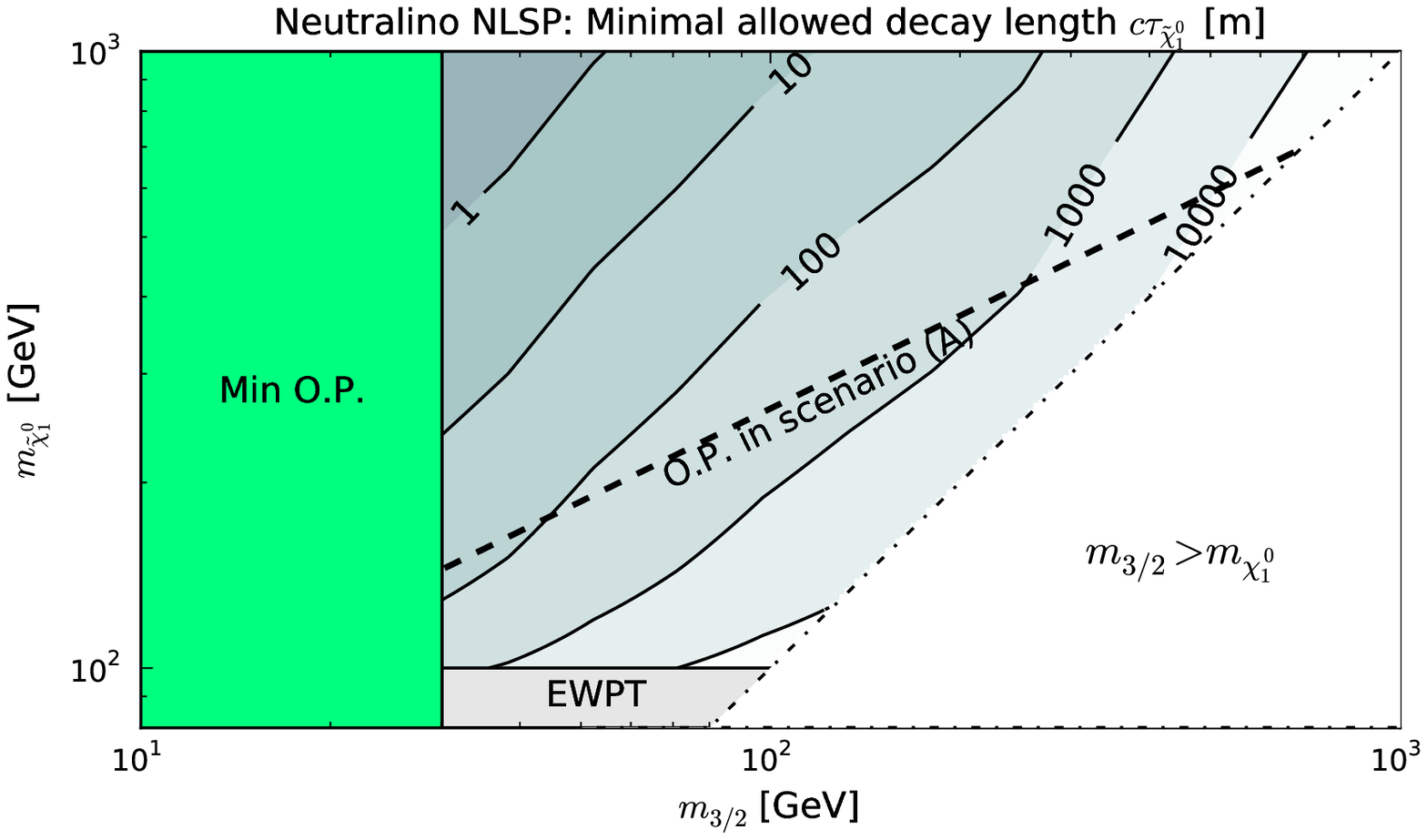}
  \end{center}
  \caption{Like Fig.~\ref{fig:Staudecaylength}, but for a neutralino NLSP.}
  \label{fig:Neutralinodecaylength}
\end{figure}

Using the gaugino mass relation Eq.~\eqref{eq:gauginomasses}, lower bounds on
the neutralino decay length $c\tau_{\tilde \chi_1^0}$ can be derived from the
partial gravitino decay width. Our results are summarized in
Fig.~\ref{fig:Neutralinodecaylength} considering the cluster lifetime
limits. For the parameter space allowed by EWPT
and overproduction bounds, we obtain minimal decay lengths ${\cal
O}(100\,\text{m}-100$\,km), which are in the range of detectability of the
LHC~\cite{Ishiwata:2008tp,Bobrovskyi:2011vx}.  Decay lengths as small as
$c\tau_{\tilde \chi_1^0} \simeq 60$\,m are allowed for $\mG \simeq 30$\,GeV at
$\mN \simeq 140$\,GeV.

\section{Conclusions}
Galaxy clusters are the most massive collapsed objects in the Universe, and
very interesting targets for the indirect search for dark matter. Using three
years of Fermi LAT data, we studied the gamma-ray emission from eight of the
most promising galaxy clusters and searched for signatures for dark matter
decay or annihilation. We analyzed the clusters individually as well as in a
combined likelihood approach. We took into account uncertainties in the
cluster masses as determined by ROSAT PSPC X-ray observations and modeled the
dark matter signals as extended sources. Our main results are:
\begin{itemize}
  \item In none of the eight galaxy clusters listed in Tab.~\ref{tab:clusters}
    a significant gamma-ray emission that could be attributed to dark matter
    decay or annihilation was found. We derived limits at the $95\%$ C.L.~on
    the dark matter lifetime and on the annihilation cross-section, from each
    cluster individually as well as in a combined likelihood approach.  Our
    results are shown in the Figs.~\ref{fig:limitsBBAR}, \ref{fig:limitsMUMU},
    \ref{fig:limitsSTACKED} and \ref{fig:limitsBOOSTED} for $b\bar{b}$,
    $\mu^+\mu^-$, $W^+W^-$ and $\tau^+\tau^-$ final states. In most cases the
    combined limits are at the level of the strongest individual limits; in
    any case the combined limits are more robust with respect to uncertainties
    of the cluster masses, background modeling and statistical fluctuations in
    the cluster target regions (Fig.~\ref{fig:limitsSIGMA}).
  \item The limits on the \textit{dark matter lifetime} turn out to be
    somewhat weaker than what previous results indicated~\cite{Dugger:2010ys},
    reaching up to lifetimes of $\tau_\text{DM}\simeq 4\times 10^{26}\s$ in
    case of decay into $b\bar{b}$ (Fig.~\ref{fig:limitsBBAR}). The difference
    can be partly attributed to the fact that we modeled the cluster emission
    as an extended signal rather than as a point-like source, and partly to
    the inclusion of cluster mass uncertainties.  We find that a point-source
    approximation to the signal from dark matter decay could in some cases
    strengthen the limits wrongly by a factor of a few
    (Fig.~\ref{fig:limitsPSC}). In particular, the decaying dark matter
    interpretation of the $e^\pm$ excess in terms of dark matter decaying into
    $\mu^+\mu^-$ remains only partially constraint by our galaxy cluster
    limits (Fig.~\ref{fig:limitsMUMU}).
  \item As long as prompt radiation dominates the overall gamma-ray signal,
    our cluster limits on the dark matter lifetime are sometimes weaker than
    the corresponding limits that can be conservatively inferred from
    measurements of the \textit{extragalactic gamma-ray background} (EGBG, see
    Figs.~\ref{fig:limitsBBAR} and~\ref{fig:limitsMUMU}). However, at lower
    gamma-ray energies the cluster and EGBG limits are comparable and should
    be considered as being complementary, since the systematics of the
    background subtraction are in general very different.
  \item In case of an \textit{annihilation signal} coming from the smooth
    component of the cluster dark matter halo alone, while neglecting the
    contributions from dark matter substructures, previously presented
    limits~\cite{Ackermann:2010rg} are improved by up to a factor of two
    (Figs.~\ref{fig:limitsBBAR}, \ref{fig:limitsMUMU} and
    \ref{fig:limitsSTACKED}). In case of annihilation into $b\bar{b}$ final
    states, we obtain limits down to $5\times10^{-25}\cm^3\s^{-1}$ for a
    $10\GeV$ WIMP, and we confirm that the point-source approximation to the
    dark matter emission is valid when the effect of dark matter substructures
    are neglected. However, as we discuss for the case of the Coma cluster,
    for such a point-like emission intra-cluster magnetic fields could reduce
    the unboosted annihilation signal by an order one factor when the signal
    is dominated by photons from ICS radiation.
  \item When contributions to the annihilation signal from \textit{dark matter
    substructures} are taken into account, the dark matter signal must be
    modeled as an extended source (\cp Fig.~\ref{fig:profile}). Adopting the
    optimistic scenario for signal boosts due to substructures from
    Ref.~\cite{Pinzke:2011ek}, we find that the limits could strengthen by a
    factor of a few hundred if substructures with masses down to
    $M_\text{lim}=10^{-6}M_\odot$ are included. In this case, the limits on
    $b\bar{b}$ would start to reach the thermal cross-section
    $3\times10^{-26}\cm^3\s^{-1}$ and exclude dark matter masses below
    $150\GeV$ (Fig.~\ref{fig:limitsBOOSTED}).
\end{itemize}

As a direct application of our results, we derived limits on the
\textit{decaying gravitino} dark matter scenario, both from galaxy clusters
observations and from the EGBG. We find that the cluster limits on the
gravitino lifetime is $\tau_{3/2} \gtrsim 1$--$2\times10^{26}$\,s for
gravitino masses up to $1\TeV$. The cluster constraint becomes stronger than
the gamma-ray line limit at gravitino masses of $m_{3/2} \gtrsim 250$\,GeV.
As a result, we found limits on the $R$-parity breaking parameter $\zeta$ of
the order of ${\cal O}(10^{-10})$ for gravitino masses 250\,GeV$\lesssim
m_{3/2} \lesssim$ 1\,TeV.  These constraints were used to set lower limits on
NLSP decay lengths corresponding to two different types of universal boundary
conditions at the grand unification scale of supergravity models.
Interestingly, all the implied decay lengths should be accessible at the LHC.

\section*{Acknowledgments}
We thank Hans B\"ohringer, Julien Lavalle, Luca Maccione, David Paneque,
Christopher Savage, Pat Scott, Georg Weiglein and Stephan Zimmer for useful
discussions.  X.H.~was supported by the exchange program between the
Max-Planck Society and the Chinese Academy of Science. C.W.~thanks the Kavli
Institute for Theoretical Physics, Beijing, for warm hospitality during the
final stages of this work.  This work was partly carried out at the Theory
Division of CERN in the context of the TH-Institute DMUH'11 (18-29 July 2011). 

\bibliographystyle{JHEP}
\bibliography{}

\end{document}